
\magnification=\magstep1

\parindent=15pt
\centerline{\hfill OKHEP-93-07}
\smallskip
\centerline{\bf DISCRETE TIME QUANTUM MECHANICS}
\bigskip
\bigskip
\bigskip
\centerline{Carl M. Bender}
\medskip
\centerline{Department of Physics}
\medskip
\centerline{Washington University}
\medskip
\centerline{St. Louis, MO 63130-4899}
\bigskip
\bigskip
\centerline{Lawrence R. Mead}
\medskip
\centerline{Department of Physics and Astronomy}
\medskip
\centerline{University of Southern Mississippi}
\medskip
\centerline{Hattiesburg, MS 39406-5046}
\bigskip
\bigskip
\centerline{Kimball A. Milton}
\medskip
\centerline{Department of Physics and Astronomy}
\medskip
\centerline{University of Oklahoma}
\medskip
\centerline{Norman, OK 73019-0225}
\bigskip
\bigskip
\bigskip
\bigskip
\bigskip
\centerline{\bf ABSTRACT}
\bigskip
This paper summarizes a research program that has been underway for a
decade.
The objective is to find a fast and accurate scheme for solving
quantum problems
which does not involve a Monte Carlo algorithm. We use an alternative
strategy
based on the method of finite elements. We are able to formulate
fully
consistent quantum-mechanical systems directly on a lattice in terms
of operator
difference equations. One advantage of this discretized formulation
of quantum
mechanics is that the ambiguities associated with operator ordering
are
eliminated. Furthermore, the scheme provides an easy way in which to
obtain the
energy levels of the theory numerically. A generalized version of
this
discretization scheme can be applied to quantum field theory
problems. The
difficulties normally associated with fermion doubling are
eliminated. Also,
one can incorporate local gauge invariance in the finite-element
formulation.
Results for some field theory models are summarized. In particular,
we review
the calculation of the anomaly in two-dimensional quantum
electrodynamics (the
Schwinger model). Finally, we discuss nonabelian gauge theories.
\footnote{}{PACS numbers: 02.90.+p, 11.90.+t, 11.10.-z}
\vfill \eject

\noindent{\bf I. INTRODUCTION}
\bigskip
A classical physical system is described by a differential equation
supplemented by initial conditions. The differential equation
characterizes the
time evolution of the system. In quantum physics one cannot specify
initial
conditions for an {\sl operator} differential equation because
specifying the
values of the fundamental observables (operators) at a given time
would violate
the uncertainty principle. Thus, a quantum theory is described by an
operator
differential equation (a field equation) supplemented by an
equal-time
commutation relation (ETCR). The ETCR is a constraint which must hold
at
each time; its constancy in time expresses the {\sl unitarity}
(conservation
of probability) of the quantum theory. The ETCR of quantum mechanics
replaces
the initial condition of classical mechanics.

In the study of quantum field theory there are inherent ambiguities
and
divergences associated with an operator-differential-equation
formulation
because operator-valued distributions in the continuum are so
singular that
products of such operators do not exist. It is well known that
introducing a
space-time lattice is a good way to remove these ambiguities and thus
to
regularize a continuum quantum field theory. The content of the
theory is then
contained in the continuum limit of the lattice theory.

Ordinarily, the lattice is introduced as a mathematical artifice to
make sense
of the functional integral representation of a quantum field-theory.
In this
computational scheme the lattice regularizes the functional integral
as an
infinite product of ordinary Riemann integrals. Then, the infinite
product is
approximated as a finite product of integrals which are evaluated by
Monte-Carlo
methods. This procedure is slow; doubling the computer time gives
only minimal
improvements in accuracy.

The program discussed here uses the lattice in a completely different
and more
fundamental way. We show how to formulate and construct a fully
consistent
(unitary) quantum theory on a space-time lattice. Such a theory is
defined in
terms of an operator difference equation (rather than a differential
equation)
and an ETCR that holds at equal-time lattice points. The operators in
such a
theory have none of the problems (infinities) that operators in the
continuum
have. We will see that, while there is no hope of solving operator
differential
equations, operator {\sl difference} equations can be solved exactly.
Thus, for
each lattice we obtain exact closed-form solutions for the field
operators
rather than a slowly converging sequence of Monte-Carlo
approximations.

To convert an operator differential equation to an operator
difference
equation we use the method of linear finite elements. This method is
explained
in Sec.~II of this paper. There we show explicitly that for a
one-particle
quantum system the discrete-time operator equation obeys the
constraint of
unitarity. We show how to solve the operator difference equation
exactly and how
to use the solution to obtain accurate numerical estimates of the
eigenvalues.
In Sec.~III we show that the method of finite elements resolves the
well-known
operator-ordering ambiguities one encounters in the usual Hamiltonian
formulation. In Sec.~IV we show that one can use higher-order finite
elements
to generate systems of operator difference equations. We show that
the
requirement of unitarity can be used to recover gaussian quadrature.
In Sec.~V
we apply the method of finite elements to Hamiltonian systems, such
as spin
systems, that are associated with algebras other than the Heisenberg
algebra.

The next five sections address the more difficult problem of systems
having
more than one degree of freedom. In Sec.~VI we consider the case of
two
degrees of freedom and show that on a finite-element lattice the
discrete
quantum system is unitary. Next, we examine the question of how to
apply the method of finite elements to quantum field theory problems,
which
are systems having an infinite number of degrees of freedom. In
Sec.~VII
we consider the simplest case; namely, that of scalar quantum field
theory.
Then, in Sec.~VIII we examine the Dirac equation and spinor field
theories. We
show that the method of finite elements resolves the fermion-doubling
problem,
a generic difficulty usually encountered whenever one attempts to
discretize the
Dirac equation. In the last two sections we use the method of finite
elements to
solve quantum field theories exhibiting local gauge invariance. In
Sec.~IX we
examine quantum electrodynamics, a quantum field theory that has an
abelian
gauge invariance. We apply our analysis to solve the Schwinger model,
massless quantum electrodynamics in two dimensions, and obtain the
anomaly.
In Sec.~X we examine quantum field theories having local nonabelian
gauge
invariance.
\vfill \eject

\noindent{\bf II. QUANTUM MECHANICS WITH ONE DEGREE OF FREEDOM}
\bigskip
\noindent{\bf A. The Method of Finite Elements}
\medskip
The method of finite elements$^1$ is a technique for solving partial
differential equations that is well known to applied mathematicians.
The method
consists of four steps. We are given a classical partial differential
equation
$L \phi = 0$, which is to be solved on a region R subject to boundary
conditions
given on the boundary $\partial R$. We first decompose $R$ into a set
of
nonoverlapping patches, called finite elements, which completely
cover $R$. For
classical (not quantum) differential equations the patches may have
arbitrary
sizes and shapes. On each patch we approximate the solution $\phi$ to
the
partial differential equation as a polynomial. The degree of this
polynomial is
chosen to suit the conditions of the problem. Second, at the
boundaries of
contiguous patches continuity is imposed (and sometimes continuity of
higher
derivatives). Third, on patches that are adjacent to $\partial R$ we
impose the
boundary conditions. Fourth, we impose the differential equation $L
\phi = 0$ at
one point (or more than one) on each patch. Conditions two, three,
and four give
a system of algebraic equations satisfied by the coefficients of the
polynomials. Solving this system gives a good approximation to the
solution
$\phi$.

We illustrate this procedure by solving a simple classical ordinary
differential equation problem:
$$y'(x) = y(x), \quad y(0)=1.\eqno\hbox{(2.1)}$$
Show that
$$y(1) = e = 2.71828\ldots .\eqno\hbox{(2.2)}$$

We begin by using just one linear finite element: $y=ax+b$, where
$0\leq x\leq
1$. The initial condition gives one algebraic equation.
$$y(0) = 1 \quad\quad\Rightarrow \quad\quad b=1.\eqno\hbox{(2.3)}$$
We must impose the differential equation at one point $x_0$ on the
interval
$0\leq x_0 \leq 1$; however, the choice of $x_0$ remains ambiguous.
Later we
will see that unitarity in quantum mechanics removes this ambiguity
and uniquely
selects $x_0 = 1/2$. For now we simply choose $x_0 = 1/2$ and
proceed:
$$y'(1/2) = y(1/2) \quad\quad\Rightarrow\quad\quad a = {a\over 2} +
b.
\eqno\hbox{(2.4)}$$
Solving (2.3) and (2.4) for $a$ and $b$ gives $y(x) = 2x+1$, so that
$y(1)=3$,
which is a good result that already differs from the exact answer by
only 10\%.

For the case of two linear finite elements $y_1 = at+b$, $y_2 =
ct+d$, where $t$
is a local variable that ranges from $0$ to $1/2$, the initial
condition gives
$$y_1(0) = 1 \quad\quad\Rightarrow\quad\quad b=1.\eqno\hbox{(2.5)}$$
Continuity at $x = 1/2$ gives
$$y_1 (1/2) = y_2 (0) \quad\quad\Rightarrow\quad\quad {a\over 2} + b
=
d\eqno\hbox{(2.6)}$$
and imposing the differential equation at the center of each finite
element
$x =1/4$ and $x=3/4$ gives
$$y'_1 (1/4) = y_1 (1/4) \quad\quad\Rightarrow\quad\quad a={a\over 4}
+ b,$$
$$y'_2 (1/4) = y_2 (1/4) \quad\quad\Rightarrow\quad\quad c={c\over
4}+d.
\eqno\hbox{(2.7)}$$

Simultaneous solution of Eqns.~(2.5--7) gives an excellent result for
y(1):
$$y(1) = y_2 (1/2) = {25\over 9} = 2.778\ldots ,\eqno\hbox{(2.8)}$$
which differs from the correct answer by 2\%.

In general, for $N$ finite elements the exact result for the
approximate value
of $y(1)$ is
$$y(1) = \left ( {{2N+1} \over {2N-1}}\right ) ^N
,\eqno\hbox{(2.9)}$$
which for large $N$ approximates $e$ with a very small relative error
of
$1/(12N^2)$.
\bigskip
\noindent{\bf B. The Equations of Quantum Mechanics}
\medskip
Applying the technique of finite elements to quantum problems is much
more
interesting because the polynomial coefficients are operators.
Consider the
simple quantum-mechanical Hamiltonian
$$H = {1\over 2}p^2 + V(q),\eqno\hbox{(2.10)}$$
for which Hamilton's equations are
$$\dot q = p, \quad \dot p = -V'(q).\eqno\hbox{(2.11)}$$
The system (2.11) constitutes a time-evolution problem for the
operators $p(t)$
and $q(t)$. The analog of the classical initial condition is an
operator
constraint in the form of an ETCR
$$[q(t), p(t)] = i .\eqno\hbox{(2.12)}$$
If (2.12) is imposed at $t=0$ then, by virtue of (2.11), it holds for
all $t$.

To solve (2.11) on the interval $[0,T]$ we introduce a lattice of $N$
linear
finite elements. On each finite element $t$ ranges from $0$ to $h$
and $Nh=T$.
Let $q_n$, $(n=0,~1,~2,\ldots,N)$ be the approximate value of
$q(nh)$. Let us
examine the $n$th finite element, where $p(t)$ and $q(t)$ are
approximated by
the linear polynomials
$$p(t) = (1-t/h) p_{n-1} + (t/h) p_n ,$$
$$q(t) = (1- t/h) q_{n-1} + (t/h) q_n ,\eqno\hbox{(2.13)}$$
with $0 \leq t \leq h$. Substituting (2.13) into (2.11) and
evaluating at the
center of the finite element $t=h/2$, we obtain a pair of algebraic
equations
relating the operators $p_{n-1}$, $q_{n-1}$, $p_n$, and $q_n$:
$${{q_n-q_{n-1}}\over h} = {{p_n+p_{n-1}}\over
2},\eqno\hbox{(2.14a)}$$
$${{p_n-p_{n-1}}\over h} = -V'\left ( {{q_n+q_{n-1}}\over 2} \right
).\eqno
\hbox{(2.14b)}$$

The ETCR (2.12) at the lattice point $nh$ reads
$$[q_n,p_n] = i. \eqno\hbox{(2.15)}$$
It is not obvious that (2.14) and (2.15) are consistent. To prove
consistency we
argue as follows: Equation (2.14a) implies that
$$[q_n - q_{n-1}, p_n + p_{n-1}] = 0 \eqno\hbox{(2.16a)}$$
and (2.14b) implies that
$$[q_n + q_{n-1}, p_n - p_{n-1}] = 0.\eqno\hbox{(2.16b)}$$
Expanding and adding together the commutators in (2.16) gives the
result
$$[q_n , p_n ] - [q_{n-1}, p_{n-1}] = 0. \eqno\hbox{(2.17)}$$
Thus, if $[q_0,p_0] = i$ initially, then (2.15) holds for all values
of $n$. The
proof of the persistence of the ETCR's holds if and only if both
differential
equations (2.11) are imposed on finite elements at $t=h/2$. {\sl At
every other
point on the finite element (2.15) ceases to be true.}$^2$ Thus,
quantum-mechanical unitarity (persistence of the ETCR's) removes a
basic
ambiguity that occurs in the numerical solution of classical
differential
equations; namely, where on the finite element to impose the
differential
equation.
\bigskip
\noindent{\bf C. Solution of the Operator Equations}
\medskip
We have proved the consistency of (2.14), but we must now solve these
difference
equations. To do so we use (2.14a) to eliminate $p_n$ from (2.14b)
which now
becomes
$${4\over {h^2}}q_{n-1}+{2\over h} p_{n-1}=V'\left (
{{q_n+q_{n-1}}\over 2}
\right ) +{4\over{h^2}}\left ( {q_n+q_{n-1}\over 2}\right)
.\eqno\hbox{(2.18)}$$
If we let $x=(q_n+q_{n-1})/2$, $y=4q_{n-1}/h^2+2p_{n-1}/h$, and
$g(x)=V'(x)+
4x/h^2$, then (2.18) becomes
$$y = g(x).\eqno\hbox{(2.19)}$$

While $x$ and $y$ are operators, (2.19) implies that they commute and
thus
(2.19) can be treated as if it were a $c$-number equation. Its exact
solution is
$x= g^{-1}(y)$ so
$$q_n  = - q_{n-1} + 2 g^{-1} \left ( {2\over h} p_{n-1} + {4\over
{h^2}}
q_{n-1} \right ) ,$$
$$p_n  = - p_{n-1} - {4\over h} q_{n-1} + {4\over h} g^{-1} \left (
{2\over h} p_{n-1} + {4\over {h^2}} q_{n-1} \right )
.\eqno\hbox{(2.20)}$$
This result shows that the exact operator solution after $N$ time
steps to the
lattice quantum theory in (2.14--15) is a continued (nested)
function.$^3$

The unitarity of the lattice theory can also be demonstrated
explicitly because
the transfer (lattice time evolution) operator $U$ can be expressed
in closed
form:
$$q_{n+1} = U q_n U^{-1}, \quad p_{n+1} = U p_n U^{-1},$$
where
$$U = e^{i h p_n^2 /4} e^{ihA(q_n )} e^{ih p_n^2
/4}\eqno\hbox{(2.21a)}$$
with
$$A(x) = {2\over {h^2}} \left [ x-{4\over {h^2}}g^{-1} (x) \right ]
^2 +
V \left [ {4\over {h^2}} g^{-1} (x) \right ].\eqno\hbox{(2.21b)}$$

It is interesting that while the solution in (2.20) and the transfer
operator
$U$ involves the function $g^{-1}$, matrix elements of these
operators only
involve $g$. For example, if we define Fock states $|n\rangle$ at the
initial
time by
$$p_0 = {1\over {i\gamma \sqrt{2}}} (a- a^{\dagger} ) ~~{\rm and}~~
q_0 = {{\gamma}\over {\sqrt{2}}} (a+ a^{\dagger} ) , $$
where
$$a |n\rangle =\sqrt{n} |n-1\rangle ~~{\rm and}~~ a^{\dagger}
|n\rangle =\sqrt
{n+1} |n+1\rangle $$
then,$^4$
$$\langle m|q_1 |n\rangle = -{{\gamma}\over{\sqrt{2}}}
(\sqrt{n}\delta_{m,n-1}+
\sqrt{m} \delta_{n,m-1} ) $$
$$ + { {e^{i\theta (n-m)}}\over {R \sqrt{\pi 2^{m+n}n!m!}}}
\int_{-\infty}^{\infty} dz\; z e^{-g^2(z)/(4R^2)} g'(z)
H_n\left ( {{g(z)}\over{2R}}\right ) H_m\left ( {{g(z)}\over {2R}}
\right ),$$
where $R^2 =4\gamma^2 h^{-4} + h^{-2} \gamma^{-2}$,
$\cos\theta=2\gamma/(Rh^2)$,
and $H_n$ is the $n$th Hermite polynomial.
\bigskip
\noindent{\bf D. Energy Eigenvalues}
\medskip
It is easy to compute energy levels of quantum systems once the
operator
equations have been solved. There are two ways to carry out such a
calculation.
The quick and approximate method makes use of the {\sl
one}-finite-element
solution to the operator difference equations. The techniques for
performing
this calculation are given elsewhere.$^5$ Here are some numerical
results: For
the case of the harmonic oscillator, the energy gap $\omega =E_1-E_0$
comes out
exactly. For the anharmonic oscillator, where $V(q)=\lambda q^4 /4$,
the exact
value of $\omega$ is $1.08845\ldots\lambda^{1/3}$. The
one-linear-finite-element
equations predict $1.14471\lambda^{1/3}$ (5.2\% relative error); the
one-quadratic-finite-element equations (see Sec.~IV) predict
$1.08225\lambda^{1/3}$ (-.57\% relative error).

It is also possible to determine {\sl all} energy differences
simultaneously and
to arbitrary accuracy by taking large numbers of finite elements. The
procedure
consists of using the method of finite elements to calculate the
matrix
element $A_n = \langle 0 | q_n | 1 \rangle$ of the operator $q_n$.
The
sequence of numbers $A_n$ is a discrete time sequence. We can then
compute
the discrete Fourier transform $\tilde A_m$ of this sequence and
search for
peaks in this distribution. These peaks correspond to energy
differences in
the spectrum of the theory. The detailed procedure for this
calculation is
given elsewhere.$^6$

Figure 1 shows the results of a short computer calculation using
$1000$ finite
elements for the anharmonic oscillator. The results in Fig.~1 are
summarized
numerically in Table 1.$^6$
\vskip 13cm
\noindent
{\bf Figure 1.}~~A semilog plot of $|\tilde A_m |^2$
versus $m$ for the anharmonic oscillator $V(q) = 0.885 q^4$. The
spikes give
extremely accurate approximations to energy differences $E_j - E_k$
of the
exact spectrum. To read off the predicted energy differences, we note
that
one unit on the horizontal scale corresponds to an energy increment
of
$\Delta = 2\pi/[(N+1)h]$. Energy differences are measured from both
the left
axis and the right boundary (see Ref.~6).
\eject
$$\vbox{\halign{\hfil #\hfil\quad &\hfil #\hfil \quad &\hfil #\hfil
\quad &
\hfil #\hfil \cr
Energy difference & Exact & Approximate & Relative error \cr
\noalign{\smallskip\hrule\bigskip}
$E_1 - E_0$ & 1.728 & 1.674 & 3.1\% \cr
\noalign{\medskip}
$E_2 - E_1$ & 2.142 & 2.218 & -3.5\% \cr
\noalign{\medskip}
$E_3 - E_2$ & 2.537 & 2.595 & -2.3\% \cr
\noalign{\medskip}
$E_4 - E_3$ & 2.790 & 2.846 & -2.0\% \cr
\noalign{\medskip}
$E_5 - E_4$ & 3.000 & 3.097 & -3.2\% \cr
\noalign{\medskip}
$E_6 - E_5$ & 3.210 & 3.306 & -3.0\% \cr
\noalign{\medskip}
$E_3 - E_0$ & 6.407 & 6.487 & -1.2\% \cr
\noalign{\medskip}
$E_4 - E_1$ & 7.469 & 7.659 & -2.5\% \cr
\noalign{\medskip}
$E_5 - E_2$ & 8.327 & 8.454 & -1.5\% \cr
\noalign{\medskip}
$E_6 - E_3$ & 9.000 & 9.165 & -1.8\% \cr
\noalign{\medskip}
$E_5 - E_0$ & 12.20 & 12.39 & -1.6\% \cr
\noalign{\medskip}
$E_6 - E_1$ & 13.68 & 13.98 & -2.2\% \cr
\noalign{\medskip}
}}$$
\bigskip
\noindent
{\bf Table 1.}~~Comparison between exact eigenvalue differences for
the
anharmonic oscillator $V(q) = g q^4$, with $g = 0.885$ and the
approximate
eigenvalue differences obtained in the following manner. We find the
integer
values of $m$ for which $|\tilde A(m)|^2$ is a local maximum. The
energy
difference $E_j - E_k$, for some $j$ and $k$, is then predicted to be
$2\pi m
/ [(N+1)h]$. This procedure gives relative errors of order 1--3\% as
the table
shows. However, this procedure can be drastically refined by using by
using
nearby values of $|\tilde A (m)|^2$ to interpolate the precise
noninteger value
of the location of the maximum.
\bigskip
Computations involving many finite elements also give extremely
accurate results
in problems involving tunneling.$^7$ Rather than solving the
time-dependent
Schr{\"o}dinger equation we can obtain quantum tunneling results
directly by
following the time evolution of the matrix elements of operators on a
discrete-time lattice.
\vfill \eject

\noindent{\bf III. GENERAL HAMILTONIAN SYSTEMS AND OPERATOR ORDERING}
\bigskip
The Hamiltonian $H$ in (2.10) is not the most general one-particle
quantum-mechanical Hamiltonian. However, the more general
one-particle
Hamiltonian, $H(p,q)$, suffers from an ambiguity that is not present
in $H$
in (2.10); namely, the well-known problem of operator ordering. There
is no such ordering problem in classical mechanics. However,
quantizing a given
classical-mechanical Hamiltonian $H(p,q)$ is an ambiguous procedure.
The
following simple example illustrates why this is so. Consider the
classical
Hamiltonian $H(p,q)=p^2 q^2$. The corresponding quantum-mechanical
Hamiltonian could consist of any of the following Hermitian
operators:
$${1\over 2} (p^2 q^2 + q^2 p^2),\quad {1\over 2} (pqpq + qpqp),\quad
p q^2 p,
\quad q p^2 q.$$
It could even be a linear combination of these operators. We show in
this
section that if we take the lattice as fundamental and use the method
of finite
elements, the above operator ordering ambiguity is completely
eliminated.$^8$

With $H(p,q)$ an arbitrary function of the operators $p$ and $q$ the
operator
differential equations of motion in the continuum are
$$\dot q (t) = {{\partial H}\over {\partial p}} = -i [q,H],$$
$$\dot p (t) = - {{\partial H}\over {\partial q}} = -i
[p,H].\eqno\hbox{(3.1)}$$

Let us address the problem of converting this system of equations in
the
continuum into a system of {\sl unitary} operator difference
equations on a time
lattice. Recall that by unitary we mean that the difference equations
exactly
preserve the equal-time commutation relation
$$[q(t), p(t)] = i\eqno\hbox{(3.2)}$$
at each time step. We show that if the method of finite elements is
used to
construct the operator difference equations then the ordering of the
operators $p$ and $q$ is uniquely determined by the unitarity
requirement.

The finite-element method consists of making the replacements
$$\dot q(t) \rightarrow (q_n - q_{n-1})/h,$$
$$\dot p(t) \rightarrow (p_n - p_{n-1})/h,$$
$$q(t) \rightarrow (q_n + q_{n-1})/2,$$
$$p(t) \rightarrow (p_n + p_{n-1})/2,\eqno\hbox{(3.3)}$$
in (3.1). Thus, on the lattice, the differential equations (3.1)
become
$$\dot Q= {{\partial}\over {\partial P}} H(P,Q), \quad
\dot P=  - {{\partial}\over {\partial Q}} H(P,Q), \eqno\hbox{(3.4)}$$
where we have used the notation
$$\dot Q \equiv (q_n - q_{n-1})/h,$$
$$\dot P \equiv (p_n - p_{n-1})/h,$$
$$Q \equiv (q_n + q_{n-1})/2,$$
$$P \equiv (p_n + p_{n-1})/2. \eqno\hbox{(3.5)}$$

To establish unitarity as in Sec.~II one must prove that
$$[q_n,p_n] = [q_{n-1},p_{n-1}],\quad
(n=0,~1,~2,~\dots).\eqno\hbox{(3.6)}$$
We can establish (3.6) if we can explicitly show from the operator
difference
equations (3.4) that
$$[\dot Q, P] + [Q, \dot P] = 0.\eqno\hbox{(3.7)}$$
To see this we substitute the definitions in (3.5) into (3.7) and
expand the
commutators; this calculation directly shows that (3.7) implies
(3.6). Thus,
our objective is to examine the expression
$$[{{\partial}\over {\partial P}} H(P,Q) , P]
- [Q, {{\partial}\over {\partial Q}} H(P,Q) ] \eqno\hbox{(3.8)}$$
and to show that it vanishes.

It is crucial to remark that in general (3.8) does {\sl not} vanish.
This is
because the commutator
$$\theta \equiv [Q, P] \eqno\hbox{(3.9)}$$
is not a $c$-number even though $[q(t),p(t)]$ is; the quantity
$\theta$ is an
operator because it contains the {\sl unequal}-time commutators
$[q_{n-1},p_n]$
and $[q_n,p_{n-1}]$.

To investigate (3.8) we may assume that $H$ is Hermitian and that
$H(P,Q)$ can
be expanded in a series of Hermitian terms $H_{m,n}(P,Q)$, $(m,n\geq
0)$,
which consists of a sum of monomials containing $m$ factors of $P$
and $n$
factors of $Q$. We can examine each term $H_{m,n}$ of the series
independently.
For example, $H_{2,2}$ has the form
$$H_{2,2}=aPQ^2P+bQP^2Q + c(P^2Q^2 + Q^2P^2)+d(PQPQ +
QPQP),\eqno\hbox{(3.10)}$$
where $a$, $b$, $c$, and $d$ are real constants. To illustrate our
procedure we
examine $H_{2,2}$ in detail. We compute
$$[{{\partial}\over {\partial P}} H_{2,2}, P]
- [Q, {{\partial}\over {\partial Q}} H_{2,2}]$$
$$= (2c-a-d)(\theta QP + PQ\theta) + (a-b)(Q\theta P + P\theta Q) +
(b+d-2c)(\theta PQ + QP\theta). \eqno\hbox{(3.11)}$$
Thus, unitarity requires that $2c-a-d=0$, $a-b=0$, and $b+d-2c=0$.
The solution
to these equations is $a=b$ and $d=2c-b$ where $b$ and $c$ are
arbitrary real
constants. Thus, it appears that there is a two-parameter family of
Hamiltonians
of the type $H_{2,2}$ which exhibit unitarity on the lattice. Indeed,
$H_{2.2}$ can be written in the form
$$H_{2,2}=c T_{2,2} + (c-b) G_{2,2},\eqno\hbox{(3.12)}$$
where
$$T_{2,2}(P,Q)=PQ^2P + QP^2Q + P^2Q^2 + Q^2P^2 + PQPQ +
QPQP\eqno\hbox{(3.13)}$$
and
$$G_{2,2} = PQPQ + QPQP - PQ^2P - QP^2Q.\eqno\hbox{(3.14)}$$

Thus, if we were given a {\sl continuum} Hamiltonian of the general
form
$H_{2,2}(p,q)$ we could reorder the operators $p$ and $q$ [using the
commutation relation (3.2)] to make it take the form in (3.12) before
making
the finite-element transcription (3.3). [Of course, this reordering
of operators
produces additional simpler terms of the $H_{1,1}(p,q)=a(pq+qp)$.]
However, we observe that $G_{2,2}(p,q)$ is trivial; using (3.2) we
see that
$G_{2,2}(p,q)=-1$. The above calculation shows that the requirement
of
lattice unitarity forces us to preorganize the operators $p$ and $q$
in
$H_{2,2}(p,q)$ in a {\sl unique} way; namely, the totally symmetric
sum ($T$
form) in (3.13). For example, if we are given the Hamiltonian
$H_{2,2}=5qp^2q$,
this Hamiltonian must be (uniquely) reordered by using (3.2) as
$$H_{2,2}(p,q) = {5\over 6} T_{2,2}(p,q) + {5\over
2}\eqno\hbox{(3.15)}$$
before going onto the lattice by use of (3.3).

What is remarkable is that given {\sl any} Hamiltonian $H_{m,n}(p,q)$
carrying
out the above procedure shows that there is always a {\sl unique}
form which is
necessary and sufficient in order that the equal-time commutators be
preserved as in (3.6). In particular, we must rewrite
$$H_{m,n}(p,q)=\alpha T_{m,n}(p,q) +
H_{m-2,n-2}(p,q),\eqno\hbox{(3.16)}$$
where $T_{m,n}$ is the totally symmetric sum ($T$ form) of all
possible
monomials containing $m$ factors of $p$ and $n$ factors of $q$. This
process
is then iterated until $H_{m,n}$ is a descending sum of totally
symmetric parts:
$$H_{m,n}(p,q)=\alpha T_{m,n}(p,q) + \beta T_{m-2,n-2}(p,q) + \dots
{}~.
\eqno\hbox{(3.17)}$$

To verify this assertion we must use the fact that derivatives leave
the $T$
form intact. In fact we have the identities
$${{\partial}\over {\partial P}} T_{m,n}(P,Q)= (m+n)
T_{m-1,n}(P,Q),$$
$${{\partial}\over {\partial Q}} T_{m,n}(P,Q)= (m+n) T_{m,n-1}(P,Q).
\eqno\hbox{(3.18)}$$
In addition, we observe that commutators maintain the totally
symmetric form
$$[Q, T_{m,n}(P,Q)] = T_{m-1,n,1}(P,Q,\theta),$$
$$[T_{m,n}(P,Q), P] = T_{m,n-1,1}(P,Q,\theta),\eqno\hbox{(3.19)}$$
where $T_{m,n,1}(P,Q,\theta)$ is the totally symmetric sum of all
monomials
having $m$ factors of $P$, $n$ factors of $Q$, and one factor of
$\theta$.
Using (3.18) and (3.19) it is easy to verify that the expression in
(3.8)
vanishes when $H(P,Q)$ is in $T$ form.

This ordering procedure applies to all Hamiltonians $H(p,q)$ which
are
polynomials in the variables $p$ and $q$. However, if $H$ is a
nonpolynomial
function the ordering problem is much more challenging. For example,
consider a class of Hamiltonians of the form
$$H(p,q) = H(T_{1,1}) = H(pq + qp).\eqno\hbox{(3.20)}$$
To order the operators of this Hamiltonian we introduce a
little-known set
of orthonormal polynomials $S_n(x)$ called continuous Hahn
polynomials.$^{9,10,11}$
These polynomials emerge from the simple observation that $T_{n,n}$
is a
polynomial function of $T_{1,1}$; the defining equation for $S_n(x)$
is
therefore$^{12,13}$
$$S_n (T_{1,1}) \equiv T_{n,n}/(2n-1)!!. \eqno\hbox{(3.21)}$$
The first few polynomials $S_n(x)$ are
$$S_0 (x) = 1,$$
$$S_1 (x) = x,$$
$$S_2 (x) = {1\over 2} (x^2 - 1),$$
$$S_3 (x) = {1\over 6} (x^3 - 5x ),$$
$$S_4 (x) = {1\over 24} (x^4 - 14 x^2 + 9),$$
$$S_5 (x) = {1\over 120} (x^5 - 30 x^3 + 89x),$$
$$S_6 (x) = {1\over 720} (x^6 -55 x^4 + 439 x^2 - 225).$$

These polynomials have the following properties:$^{10}$
\medskip
(i) The generating function $G(t)$ is
$$G(t) = {{e^{{\rm arctan}t}}\over { (1+t^2)^{1/2}}} =
\sum_{n=0}^{\infty}
S_n (x) t^n.\eqno\hbox{(3.22)}$$

(ii) The orthonormality condition is
$$\int_{-\infty}^{\infty}dx\;
w(x)S_m(x)S_n(x)=\delta_{mn},\eqno\hbox{(3.23)}$$
where the weight function $w(x)$ is given by$^{14}$
$$w(x) = [2 \cosh (\pi x/2)]^{-1}.\eqno\hbox{(3.24)}$$

(iii) A recursion relation satisfied by $S_n(x)$ is
$$n S_n (x) = x S_{n-1} (x) - (n-1) S_{n-2} (x).\eqno\hbox{(3.25)}$$
The polynomials do not satisfy a second-order differential equation
but they
do obey the second-order functional difference eigenvalue equation
$$(1-ix)S_n(x+2i) + (1+ix)S_n (x-2i) = (4n+2) S_n
(x).\eqno\hbox{(3.26)}$$

Now we return to the problem of ordering the operator $H$ in (3.20).
Using
the completeness of $S_n(x)$ we expand $H(x)$ as a series in
$S_n(x)$:
$$H(x) = \sum_{n=0}^{\infty} a_n S_n (x),\eqno\hbox{(3.27)}$$
where
$$a_n = \int_{-\infty}^{\infty}dx\; w(x)H(x)
S_n(x).\eqno\hbox{(3.28)}$$
Thus, from (3.21) we have
$$H(T_{1,1}) = \sum_{n=0}^{\infty} a_n
T_{n,n}/(2n-1)!!.\eqno\hbox{(3.29)}$$
We have therefore represented $H(pq+qp)$ as an infinite sum of
operators in $T$
form. In the form (3.29) $H$ is directly amenable to lattice
transcription and
the resulting operator difference equations automatically preserve
unitarity.

As an example, consider the Hamiltonian $H=e^{c(pq+qp)}$, where $c$
is a
constant. For the exponential function, the integral in (3.28) can be
performed
in closed form and the result is $a_n=(\tan c)^n [1+(\tan
c)^2]^{1/2}$. Thus,
$$H = [1+(\tan c)^2]^{1/2} \sum_{n=0}^{\infty} (\tan c)^n S_n
(pq+qp)$$
$$= [1+(\tan c)^2]^{1/2} \sum_{n=0}^{\infty} {{(\tan
c)^n}\over{(2n-1)!!}}
T_{n,n}.\eqno\hbox{(3.30)}$$
$H$ is now in its {\sl unique} $T$ form and therefore the resulting
Heisenberg
equations can be transcribed onto the lattice.
\vfill \eject
\noindent{\bf IV. HIGHER-ORDER FINITE ELEMENTS}
\bigskip
It is possible to generalize the linear polynomials representing
$p(t)$ and
$q(t)$ to polynomials in arbitrary degree $r$:
$$p(t) = \sum_{k=0}^r a_k (t/h)^k,\eqno\hbox{(4.1a)}$$
$$q(t) = \sum_{k=0}^r b_k (t/h)^k,\eqno\hbox{(4.1b)}$$
It is now necessary to determine $r+1$ pairs of coefficients on each
finite
element interval. The procedure for doing so is evidently ambiguous;
if we
impose the differential equations (2.11) $d$ times on the $n$th
interval,
then it is necessary to impose $r+1-d$ joining conditions
(continuity,
continuity of the first derivative, continuity of the second
derivative, and
so on) at $t=(n-1)h$. On the first interval, there are no joining
conditions at $t=0$; rather we must impose $r+1-d$ initial
conditions, in
which the values of $q(0)$ and $p(0)$, $\dot q(0)$ and $\dot p(0)$,
$\ddot q(0)$ and $\ddot p(0)$, and so on, are specified. These values
are
obtained by successively differentiating the differential equations
(2.11). We
say that as the number of joining conditions increases the
approximation becomes
{\sl stiffer}. In one extreme, the stiffest approximation, the
differential
equation is imposed once on the interval, and in the other extreme,
the floppy
approximation, the method we will use in this paper, the differential
equation
is imposed $r$ times, and we require only that the approximation be
continuous.
\bigskip
\noindent {\bf A. Failure of the Stiff Approximation}
\medskip
The stiff approximation is forbidden by quantum mechanics. For a
quantum-mechanical system with operators $p$ and $q$ the $r$th-degree
finite-element approximation is given by equations (4.1). While we
could
determine the coefficients $a_k$ on the $n$th interval from those on
the
$(n-1)$st interval, attempting to determine the coefficients on the
first
interval, even in principle, leads to an inconsistency. This is
because the
coefficients $a_k$ are operators.

We illustrate this problem by a simple example
for which $r=2$. On the first interval we represent
$$p(x) = p_0 + a_1 {x\over h} + a_2 {{x^2}\over
{h^2}}\eqno\hbox{(4.2a)}$$
and
$$q(x) = q_0 + b_1 {x\over h} + b_2 {{x^2}\over
{h^2}}.\eqno\hbox{(4.2b)}$$
For the sake of complete generality we impose the differential
equations (2.11)
at $\alpha h$ and $\beta h$, respectively, where
$0\leq\alpha\leq\beta\leq 1$
are as yet undetermined:
$${{b_1}\over h} + {{2b_2}\over h}\alpha = p_0 +a_1\alpha +
a_2\alpha^2,\eqno\hbox{(4.3a)}$$
$${{a_1}\over h} + {{2a_2}\over h}\beta = -V'(q_0 + b_1\beta +
b_2\beta^2).\eqno\hbox{(4.3b)}$$
Next, we impose the initial conditions. The condition (2.12) reads
$$[q_0, p_0] = i .\eqno\hbox{(4.4)}$$
We also impose two more commutator conditions which we obtain from
the
continuum equations at $t=0$.
$$[q(0), \dot p(0)] = [q_0,a_1]=0,\eqno\hbox{(4.5)}$$
$$[\dot q(0), p(0)] = [b_1,p_0]=0.\eqno\hbox{(4.6)}$$
There are two more commutator conditions that follow from the
equations of
motion (4.3):
$$[b_1 + 2\alpha b_2, p_0 + \alpha a_1 + \alpha^2 a_2]
=0,\eqno\hbox{(4.7)}$$
$$[a_1 + 2\beta a_2, q_0 + \beta b_1 + \beta^2 b_2]
=0.\eqno\hbox{(4.8)}$$
The five commutators (4.4--8) are kinematical in nature; they make no
reference
to the dynamical content of the theory, which is embodied in the
function $V$.

For this quantum system to be internally consistent, (4.4--6), the
analogs of
the three equal-time commutators, must hold again at $t=h$; that is
$$[q_0 + b_1 + b_2, p_0 + a_1 + a_2] =i,\eqno\hbox{(4.9)}$$
$$[q_0 + b_1 + b_2, a_1 + 2 a_2] =0,\eqno\hbox{(4.10a)}$$
$$[p_0 + a_1 + a_2, b_1 + 2 b_2] =0.\eqno\hbox{(4.10b)}$$
We can show that if (4.10) is assumed to hold, then (4.9) holds so
long as
$\alpha + \beta = 1$. However, (4.10) does not hold in general unless
$\alpha = 1$ and $\beta = 1$, which implies the failure of (4.9).

This kind of demonstration can be given for any stiff approximation
to a
quantum system. Thus, on the basis of quantum-mechanical consistency
we
reject any kind of stiff finite-element scheme in which more than a
single
initial commutator is imposed.

Also, even if a successful stiff scheme could be found, we would
prefer not
to use it because stiff schemes are not as accurate as floppy
approximations.
A maximally stiff approximation yields a relative error of $N^{-r}$
between
the exact solution and the finite-element approximation to the exact
solution.
On the other hand, for a floppy approximation, the relative error
between
these two quantities is $N^{-2r}$.
\bigskip
\noindent {\bf B. Consistency of the Floppy Approximation}
\medskip
The failure of the stiff approximation discussed the previous
subsection is not
very surprising and it is all the more remarkable that the floppy
approximation
{\sl is} successful. We begin by examining the case $r=1$, using the
notation of
the preceding subsection.
\medskip
\noindent{1. {\sl Case} $r=1$}

Imposing equations (2.11) at $t=\alpha h$ and $t=\beta h$,
respectively,
yields [here $q_1 = q(1h)$ and $p_1 = p(1h)$]
$${{q_1-q_0}\over h} = (1-\alpha) p_0 + \alpha
p_1,\eqno\hbox{(4.11)}$$
$${{p_1-p_0}\over h} = -V'[(1-\beta) q_0 + \beta
q_1],\eqno\hbox{(4.12)}$$
in the first finite element. Since $p(0) = p_0$ and $q(0)=q_0$ we
have
$$a_0 = p_0, \quad b_0 = q_0$$
and the continuity conditions at $t=h$ are
$$a_1 = p_1 - p_0, \quad b_1 = q_1 - q_0.$$
Is there a choice for $\alpha$ and $\beta$ such that (4.11) and
(4.12)
together with (2.11), the equal-time commutator at $t=0$, imply that
$[q_1,p_1]=i$? Equations (4.11) and (4.12) yield the following
commutators:
$$[q_1-q_0, (1-\alpha) p_0 + \alpha p_1] = 0,\eqno\hbox{(4.13)}$$
$$[(1-\beta) q_0 + \beta q_1, p_1 - p_0] = 0.\eqno\hbox{(4.14)}$$
Combining the three commutation relations (2.11), (4.13) and (4.14)
does
indeed yield $[q_1,p_1]=i$ provided that $\alpha$ and $\beta$ satisfy
the
constraint
$$\alpha + \beta = 1.\eqno\hbox{(4.15)}$$
Having shown consistency with quantum mechanics on the first finite
element
it follows on all finite elements by virtue of the continuity
condition on
$p(t)$ and $q(t)$ at the boundaries of adjacent finite elements,
$t=nh$.

In this section we are primarily interested in the symmetric choice
$\alpha
= \beta = 1/2$, where the equations of motion are imposed at the
midpoints
of the finite elements. Any other choice for $\alpha$ and $\beta$
breaks
time-reversal symmetry and leads to numerical approximations which
are not
as accurate as in the symmetric case.
\medskip
\noindent{2. {\sl Case} $r=2$}

Here we impose the equations of motion (2.10) twice on each finite
element. In
view of our above remarks with regard to symmetry and numerical
accuracy, we
will restrict our attention to the symmetric case where both of the
equations of
motion are imposed at the {\sl same} points $x=\alpha_1 h$ and $x
=\alpha_2 h$.

On the $n=1$ finite element we have
$$p(x) = p_0 + a_1 {x\over h} + a_2 {{x^2}\over {h^2}},$$
$$q(x) = q_0 + b_1 {x\over h} + b_2 {{x^2}\over {h^2}}.$$
Imposing (2.11) at $x=\alpha_1 h$ and $x=\alpha_2 h$ gives
$${{b_1}\over h} + {{2b_2}\over h}\alpha_1 = p_0 +a_1\alpha_1 +
a_2\alpha_1^2,$$
$${{a_1}\over h} + {{2a_2}\over h}\alpha_1 = -V'(q_0 + b_1\alpha_1 +
b_2\alpha_1^2),$$
$${{b_1}\over h} + {{2b_2}\over h}\alpha_2 = p_0 +a_1\alpha_2 +
a_2\alpha_2^2,$$
$${{a_1}\over h} + {{2a_2}\over h}\alpha_2 = -V'(q_0 + b_1\alpha_2 +
b_2\alpha_2^2). \eqno\hbox{(4.16)}$$

{}From these equations we obtain the kinematical commutators:
$$[p_0 + a_1 \alpha_1 + a_2 \alpha_1^2, b_1 + 2b_2\alpha_1]=0,$$
$$[a_1 + 2a_2\alpha_1, q_0 + b_1 \alpha_1 + b_2 \alpha_1^2] =0,$$
$$[p_0 + a_1 \alpha_2 + a_2 \alpha_2^2, b_1 + 2b_2\alpha_2]=0,$$
$$[a_1+2a_2\alpha_2,q_0 + b_1 \alpha_2 + b_2 \alpha_2^2]
=0.\eqno\hbox{(4.17)}$$
By adding these commutators we can prove that
$$[q_1,p_1] = [q_0,p_0] = i$$
if and only if $\alpha_1$ and $\alpha_2$ are given by
$$\alpha_1 = {1\over 2} - {1\over {\sqrt {12}}},\quad
\alpha_2 = {1\over 2} + {1\over {\sqrt {12}}}.\eqno\hbox{(4.18)}$$
This is the condition for quantum consistency.
\medskip
\noindent{3. {\sl Case} $r=3$}

Now we impose the equations of motion three times on each finite
element at
$x=\alpha_1 h$, $x=\alpha_2 h$, and $x=\alpha_3 h$. Taking
$$p(x) = p_0 + a_1 {x\over h} + a_2 {{x^2}\over {h^2}} + a_3
{{x^3}\over{h^3}}$$
and
$$q(x) = q_0 + b_1 {x\over h} + b_2 {{x^2}\over {h^2}}+b_3
{{x^3}\over{h^3}}$$
we obtain six equations analogous to (4.16) from which we derive six
commutator conditions analogous to (4.17). Once again we add the six
commutator conditions together. However, now the two commutators at
$x=\alpha_2 h$ are weighted by the factor $8/5$. The condition for
quantum
consistency is now
$$\alpha_1 = {1\over 2} - \sqrt{3/20},$$
$$\alpha_2 = {1\over 2},$$
$$\alpha_3 = {1\over 2} + \sqrt{3/20}.\eqno\hbox{(4.19)}$$
\medskip
\noindent{4. {\sl General Case}}

The sequence of points $\alpha$ at which the operator equations of
motion
must be imposed, $1/2$ for $r=1$, $1/2 \pm 1/\sqrt{12}$ for $r=2$,
$1/2$ and
$1/2 \pm \sqrt{3/20}$ for $r=3$, fits a well-known pattern. These
numbers are
the zeros of the $r$th Legendre polynomial $P_r(2\alpha -1)$. The
first
three such polynomials are
$$P_1(2\alpha -1) = 2\alpha -1,$$
$$P_2 (2\alpha -1)= 6\alpha^2 - 6\alpha +1,$$
$$P_3 (2\alpha -1)= (2\alpha -1)(10\alpha^2 - 10\alpha +1).$$
These zeros are the so-called Gaussian knots or nodes which are used
to
perform quadrature integration. The weighting of the commutators
necessary
to derive the consistency condition (the factor of $8/5$ mentioned
above) is
exactly the weighting used in Gaussian quadrature.$^{15}$

We conclude this section by reemphasizing that the only way to
preserve the
equal-time commutation relations is to impose the operator equations
of
motion at the Gaussian nodes. If the commutator relations are
preserved at
successive intervals of time, then the theory is unitary; that is,
there
exists a transfer operator [like that in (2.21) for linear finite
elements] that
is unitary and therefore probability is conserved. This same point
has been
observed in a totally different context by Durand,$^{16}$ who showed
that a
lattice discretization of the Schr{\"o}dinger equation preserves
orthonormality
of the wave functions only if the lattice points lie at the Gaussian
knots.
\vfill \eject
\noindent{\bf V. OTHER ALGEBRAS}
\bigskip

The successful discretization of operator equations arising from
Hamiltonian
systems associated with the Heisenberg algebra (2.12) raises an
obvious
question; namely, is it possible to discretize the operator equations
associated with other algebras? For example, consider the algebra
associated
with the rotation group $SO(3)$:
$$[X,Y]=iZ,\quad [Y,Z]=iX,\quad [Z,X]=iY.\eqno\hbox{(5.1)}$$
If we construct a Hamiltonian, $H(X,Y,Z)$, associated with this
algebra, then
the continuum equations of motion for the operator variables $X(t)$,
$Y(t)$,
and $Z(t)$ are
$$\dot X(t) = -i[X,H],$$
$$\dot Y(t) = -i[Y,H],$$
$$\dot Z(t) = -i[Z,H].\eqno\hbox{(5.2)}$$
The exact solution to the operator equations (5.2) satisfies the ETCR
in (5.1).

It is not easy to find a way to approximate such a system by a set of
operator difference equations that preserve the commutators in (5.1)
at each
time step. For example, consider the Hamiltonian spin system
$$H(X,Y,Z) = XY + YX.\eqno\hbox{(5.3)}$$
For this system, the continuum equations of motion are
$$\dot X(t) = XZ+ZX,$$
$$\dot Y(t) = -YZ-ZY,$$
$$\dot Z(t) = 2Y^2-2X^2.\eqno\hbox{(5.4)}$$
The linear finite-element prescription for discretizing continuum
equations of
motion replaces undifferentiated variables by averages and first
derivatives by
forward differences. Here we consider two adjacent lattice sites
which we label
$1$ and $2$. In terms of the discrete $X$, $Y$, and $Z$ variables the
equations
of motion (5.4) become
$${{X_2-X_1}\over h} = \left ( {{X_2+X_1}\over 2} \right )
\left ( {{Z_2+Z_1}\over 2} \right ) + \left ( {{Z_2+Z_1}\over 2}
\right )
\left ( {{X_2+X_1}\over 2} \right ),$$
$${{Y_2-Y_1}\over h} = - \left ( {{Y_2+Y_1}\over 2} \right )
\left ( {{Z_2+Z_1}\over 2} \right ) - \left ( {{Z_2+Z_1}\over 2}
\right )
\left ( {{Y_2+Y_1}\over 2} \right ),$$
$${{Z_2-Z_1}\over h} = 2 \left [ \left ( {{Y_2+Y_1}\over 2} \right
)^2
- \left ( {{X_2+X_1}\over 2} \right )^2 \right ],\eqno\hbox{(5.5)}$$
where $h$ is the lattice spacing. The statement of unitarity is that
the
SO(3) commutation relations hold at each lattice site. That is, if
$X_1$,
$Y_1$, and $Z_1$ satisfy
$$[X_1,Y_1]=iZ_1,\quad [Y_1,Z_1]=iX_1,\quad
[Z_1,Z_1]=iY_1,\eqno\hbox{(5.6)}$$
then as a consequence of (5.5), $X_2$, $Y_2$, and $Z_2$ satisfy the
same
equations.

It is easy to see that Eqs.~(5.5) violate unitarity. We merely solve
(5.5)
perturbatively in powers of the lattice spacing $h$. We seek
solutions for
$X_2$, $Y_2$, and $Z_2$ in the form
$$X_2 = X_1 +Ah + Bh^2 + Ch^3 + \ldots,$$
$$Y_2 = Y_1 +Dh + Eh^2 + Fh^3 + \ldots,$$
$$Z_2 = Z_1 +Gh + Hh^2 + Ih^3 + \ldots,\eqno\hbox{(5.7)}$$
We insert (5.7) into (5.5) and compare powers of $h$ to obtain
explicit
solutions for the operator coefficients $A$, $B$, $C$, $\dots$. The
results are
$$A = X_1 Z_1 + Z_1 X_1,$$
$$B = 2Y_1 X_1 Y_1 + 2Z_1 X_1 Z_1 - 2 X_1^3 + {3\over 2}X_1,$$
$$C = Z_1^2 X_1 Z_1 + Z_1 X_1 Z_1^2 - 4(X_1^2 Z_1 X_1 + X_1 Z_1
X_1^2)
-{3\over 4} (X_1 Z_1 + Z_1 X_1),$$
$$D = -Y_1 Z_1 - Z_1 Y_1,$$
$$E = 2Z_1 Y_1 Z_1 + 2X_1 Y_1 X_1 - 2 Y_1^3 + {3\over 2}Y_1,$$
$$F = -Z_1^2 Y_1 Z_1 + Z_1 Y_1 Z_1^2 + 4(Y_1^2 Z_1 Y_1 + Y_1 Z_1
Y_1^2)
+{3\over 4} (Y_1 Z_1 + Z_1 Y_1),$$
$$G =  2(Y_1^2 - X_1^2),$$
$$H = -4 Y_1 Z_1 Y_1 - 4 X_1 Z_1 X_1 -2 Z_1,$$
$$I = 4(X_1^4 -Y_1^4) + 6 Y_1 Z_1^2 Y_1 - 6 X_1 Z_1^2 X_1 + {5\over
2}
(Y_1^2 - X_1^2).\eqno\hbox{(5.8)}$$
We can now compute the commutator of $X_2$ with $Y_2$ to order $h^3$:
$$[X_2,Y_2] = iZ_2 + 4i(Y_1 Z_1^2 Y_1 - X_1 Z_1^2 X_1) h^3 +
i(X_1^2-Y_1^2)h^3 + {\rm O}(h^4).\eqno\hbox{(5.9)}$$
Observe that the difference between $[X_2,Y_2]$ and $iZ_2$ is not
zero but
rather of order $h^3$, showing that the lattice equations (5.5)
violate
unitarity. It is not surprising that the violation first occurs in
order
$h^3$ because we already know that for any differential equation the
expansion (5.7) agrees with the continuum result through order $h^2$,
and
that the continuum equations, of course, do not violate unitarity. We
do not
know of any simple way to avoid this violation of unitarity.

Nevertheless, there is an indirect technique for obtaining a unitary
set of
lattice equations of motion. We merely convert the spin system to an
equivalent Heisenberg system having {\sl two} degrees of freedom by
means of
the Schwinger transformation:
$$X = {1\over 4} [ (q_1-ip_1)(q_2+ip_2) + (q_2-ip_2)(q_1+ip_1) ],$$
$$Y = {i\over 4} [ - (q_1-ip_1)(q_2+ip_2) + (q_2-ip_2)(q_1+ip_1) ],$$
$$X = {1\over 4} [ (q_1-ip_1)(q_1+ip_1) - (q_2-ip_2)(q_2+ip_2) ].
\eqno\hbox{(5.10)}$$
The sorts of equations that result from this transcription will be
discussed
in the next section.
\vfill\eject
\noindent{\bf VI. SYSTEMS WITH MANY DEGREES OF FREEDOM}
\bigskip

For systems having more than one degree of freedom it is not as easy
to show that the method of finite elements is consistent with
unitarity.
We illustrate this situation with a system having two degrees of
freedom $(p,q)$ and $(\pi,\phi)$.  For the Hamiltonian
$$H={p^2\over2}+{\pi^2\over2}+V(q,\phi),\eqno(6.1)$$
Hamilton's equations in the continuum read
$$\eqalignno{\dot q&=p,&(6.2\hbox{a})\cr
\dot\phi&=\pi,&(6.2\hbox{b})\cr
\dot p&=-{\partial\over\partial q}V(q,\phi),&(6.2\hbox{c})\cr
\dot\pi&=-{\partial\over\partial\phi}V(q,\phi).&(6.2\hbox{d})\cr}$$
On the lattice, the linear finite-element transcription of these
equations of motion is
$$\eqalignno{{q_1-q_0\over h}&={p_1+p_0\over2},&(6.3\hbox{a})\cr
{\phi_1-\phi_0\over h}&={\pi_1+\pi_0\over2},&(6.3\hbox{b})\cr
{p_1-p_0\over h}&=-{\partial\over\partial q}
V\left({q_1+q_0\over2},{\phi_1+\phi_0\over2}\right),&(6.3\hbox{c})\cr
{\pi_1-\pi_0\over h}&=-{\partial\over\partial \phi}
V\left({q_1+q_0\over2},{\phi_1+\phi_0\over2}\right).&(6.3\hbox{d})\cr
}$$

Note that while there is no operator-ordering problem in (6.2c) and
(6.2d)
because $[q(t),\phi(t)]=0$, there appears to be a serious ordering
problem
in (6.3c) and (6.3d) because it is not clear that $(q_1+q_0)/2$ and
$(\phi_1+\phi_0)/2$ commute.  To resolve this problem we define
$$\alpha={2p_0\over h}+{4q_0\over h^2},\quad \beta={2\pi_0\over
h}+{4\phi_0\over h^2},\eqno(6.4)$$
and $$\sigma={q_0+q_1\over2},\quad
\tau={\phi_0+\phi_1\over2}.\eqno(6.5)$$
Now, (6.3c) and (6.3d) become
$$\eqalignno{\alpha&={\partial\over\partial\sigma}
V(\sigma,\tau)+{4\over h}
\sigma,&(6.6\hbox{a})\cr
\beta&={\partial\over\partial\tau}V(\sigma,\tau)+{4\over h}
\tau.&(6.6\hbox{b})\cr}$$
The simultaneous solution of (6.6)  has the form
$$\sigma=\sigma(\alpha,\beta),\quad
\tau=\tau(\alpha,\beta),\eqno(6.7)$$
But, $\alpha$ and $\beta$ involve only operators at the initial time,
so
$[\alpha,\beta]=0$. Thus, $\sigma$ and $\tau$ also commute, and
there is, in fact, {\it no\/} ordering problem in (6.3).

It is now necessary to verify that the ETCR's are preserved in time.
{}From the solutions
$$\eqalignno{q_1&=-q_0+2\sigma(\alpha,\beta),&(6.8\hbox{a})\cr
p_1&=-p_0-{4\over h}q_1+{4\over
h}\sigma(\alpha,\beta),&(6.8\hbox{b})\cr
\phi_1&=-\phi_0+2\tau(\alpha,\beta),&(6.8\hbox{c})\cr
\pi_1&=-\pi_0-{4\over h}\phi_1+{4\over
h}\tau(\alpha,\beta),&(6.8\hbox{d})\cr
}$$
it is necessary to verify that
$$[q_1,p_1]=[\phi_1,\pi_1]=i \eqno(6.9\hbox{a})$$
and
$$[q_1,\phi_1]=[q_1,\pi_1]=[\phi_1,p_1]=[p_1,\pi_1]=0.
\eqno(6.9\hbox{b})$$
It is easy to verify (6.9a):
$$\eqalign{[q_1,p_1]&=[q_0,p_0]+{4\over h}[q_0,
\sigma(\alpha,\beta)]
+2[p_0,\sigma(\alpha,\beta)]\cr
&=i+{4\over h}{\partial\sigma\over\partial\alpha}{2i\over h}
+2{\partial\sigma\over\partial\alpha}
\left(-{4i\over h^2}\right)\cr
&=i.\cr}\eqno(6.10)$$
However, it is harder to verify (6.9b):
$$\eqalign{[q_1,\phi_1]&=-2[q_0,\tau(\alpha,\beta)]
-2[\sigma(\alpha,\beta),
\phi_0]\cr
&=-{4i\over h}{\partial\tau\over\partial \alpha}+{4i\over
h}{\partial\sigma
\over\partial\beta}.\cr}\eqno(6.11)$$
We can show that $[q_1,\phi_1]=0$ by verifying that
$\partial\tau/\partial\alpha
=\partial\sigma/\partial\beta$ because the system defined by (6.3) is
Hamiltonian.  Explicitly, we have
$${\partial\tau\over\partial\alpha}=
{\partial\sigma\over\partial\beta}=
{-{\partial^2V\over\partial\sigma\partial\tau}
\over\left({\partial^2V\over
\partial\sigma^2}+{4\over
h^2}\right)\left({\partial^2V\over\partial\tau^2}
+{4\over
h^2}\right)-{\partial^2V\over\partial\sigma\partial\tau}}.
\eqno(6.12)$$
The other commutators are evaluated similarly.
For details and discussion of the time evolution operator see
Ref.~18.

We conclude this section with a discussion of a coupled fermion-boson
system.
Consider the system governed by the continuum Hamiltonian
$$H={p^2\over2}+S(x)+\overline\psi\psi W(x).\eqno(6.13)$$
The continuum Heisenberg equations of motion for this system are
$$\eqalignno{\dot x&=p,&(6.14\hbox{a})\cr
\dot p&=-S'(x)-\overline\psi W'(x) \psi,&(6.14\hbox{b})\cr
i\dot\psi&=W(x)\psi,&(6.14\hbox{c})\cr
-i\dot{\overline\psi}&=\overline\psi W(x).&(6.14\hbox{d})\cr}$$
To derive these equations we make use of the canonical equal-time
commutation and anticommutation relations for the dynamical variables
$$\eqalignno{[x,p]&=i,&(6.15\hbox{a})\cr
[\psi,\overline\psi]_+&=1,&(6.15\hbox{b})\cr
[x,\psi]&=[x,\overline\psi]=[p,\psi]=[p,\overline\psi]
=0,&(6.15\hbox{c})\cr
\psi^2&=\overline\psi^2=0.&(6.15\hbox{d})\cr}$$
In (6.14) we have ordered the operators in order to make the
equations of motion manifestly Hermitian, which ordering is
irrelevant in the
continuum, but not on the lattice.

If we were now to put the system (6.14) on the lattice using our
finite-element
prescription, we would find that the resulting lattice theory is not
unitary.  In anticipation of this difficulty we will replace the
function
$W'$ in (6.14b) by a function $Z$, which will be determined by the
requirement
that the system be unitary.  We will see that $Z$ satisfies an
interesting
nonlinear equation.  We will, of course, find that as the lattice
spacing
$h$ tends to zero, $Z$ approaches $W'$.

We take, then, for our lattice difference equation
$$\eqalignno{{x_1-x_0\over h}&={p_1+p_0\over2},&(6.16\hbox{a})\cr
{p_1-p_0\over h}&=-S'(\sigma)-\overline\phi
Z(\sigma)\phi,&(6.16\hbox{b})\cr
i{\psi_1-\psi_0\over h}&=W(\sigma)\phi,&(6.16\hbox{c})\cr
-i{\overline\psi_1-\overline\psi_0\over h}&=
\overline\phi W(\sigma),&(6.16\hbox{d})\cr}$$
where
$$\eqalignno{\sigma&\equiv{x_1+x_0\over2},&(6.17\hbox{a})\cr
\phi&\equiv{\psi_1+\psi_0\over2}.&(6.17\hbox{b})\cr}$$

We will now simply quote the results found in Ref.~18. There is no
problem
with unitarity in the pure fermion sector of the theory, because it
is
easy to show that
$$\psi_1={1-ihW(\sigma)/2\over1+ihW(\sigma)/2}\psi_0.\eqno(6.18)$$
Similarly, there is no problem in the pure boson sector.  The
constraint
comes when we examine the mixed commutator $[x_1,\psi_1]$.
Requiring this to vanish leads to the following equation for $Z$:
$$\eqalign{Z(x)&={W'(x)\over1+h^2S''(x)/4}\cr
&\quad+[1+h^2W^2(x)/4]\left(
S'(x)-S'\left(x-{h^2W'(x)/4\over(1+h^2W^2(x)/4)
(1+h^2S''(x)/4)}\right)\right).}\eqno(6.19)$$
For small lattice spacing this approaches
$$Z(x)=W'(x)-{S'''(x)(W'(x))^2h^4\over32}+O(h^6).\eqno(6.20)$$
Notice that if $S$ is a polynomial of degree two, $Z$ is exactly
$W'$.
This suggests that electrodynamics will not present subtleties when
analyzed using finite elements, while supersymmetry, for example, may
require more care.
\vfill\eject
\noindent{\bf VII.~SCALAR QUANTUM FIELD THEORY}
\bigskip

It is straightforward to generalize the above discussion to quantum
field theory.  In this section we will apply the method of finite
elements to self-interacting, two-dimensional scalar field theory,
and in particular, calculate the mass renormalization for the
$(\phi^{2N})_2$
and the sine-Gordon field theories.

Consider the Hamiltonian
$$H={1\over2}\phi_t^2+{1\over2}\phi_x^2+V(\phi).\eqno{(7.1)}$$
This gives rise to the operator Klein-Gordon equation
$$\phi_{tt}-\phi_{xx}=f(\phi),\eqno{\hbox{(7.2a)}}$$
where
$$f(\phi)=-V'(\phi).\eqno{\hbox{(7.2b)}}$$
Because we will be using linear finite elements, we rewrite (7.2a)
as a system of first-order equations:
$$\eqalignno{\phi_t&=\pi,&\hbox{(7.3a)}\cr
\phi_x&=\Gamma,&\hbox{(7.3b)}\cr
\pi_t-\Gamma_x&=f(\phi).&\hbox{(7.3c)}\cr}$$
We introduce a rectangular finite-element lattice with the
time-lattice
spacing being $h$ and the space-lattice spacing being $\Delta$.
The spatial extent of the lattice is $L$.
If we approximate the field in the finite element by a polynomial
linear
in $x$ and $t$,
$$\eqalignno{\phi(x,t)&=\left(1-{t\over h}\right)\left(1-{x\over
\Delta}\right) \phi_{m-1,n-1}
+\left(1-{t\over h}\right){x\over \Delta}\phi_{m,n-1}\cr
&\quad+{t\over h}
\left(1-{x\over \Delta}\right)\phi_{m-1,n}+{t\over h}{x\over
\Delta}\phi_{m,n},&(7.4)}$$
and impose the equations of motion (7.3) at the center of the finite
element,
we obtain the system of difference equations
$$\eqalignno{{1\over2h}(\phi_{m,n+1}&+\phi_{m+1,n+1}
-\phi_{m,n}-\phi_{m+1,n})\cr
&={1\over4}(\pi_{m+1,n+1}+\pi_{m,n+1}
+\pi_{m+1,n}+\pi_{m,n}),&(7.5\hbox{a})\cr
{1\over2\Delta}(\phi_{m+1,n+1}
&+\phi_{m+1,n}-\phi_{m,n+1}-\phi_{m,n})\cr
&={1\over4}(\Gamma_{m+1,n+1}+
\Gamma_{m,n+1}+\Gamma_{m+1,n}+\Gamma_{m,n}),&(7.5\hbox{b})\cr
{1\over2h}(\pi_{m,n+1}+\pi_{m+1,n+1}&-\pi_{m,n}-\pi_{m+1,n})
-{1\over2\Delta}
(\Gamma_{m+1,n+1}+\Gamma_{m+1,n}-\Gamma_{m,n+1}-\Gamma_{m,n})\cr
&=f\left({\phi_{m+1,n+1}+\phi_{m,n+1}
+\phi_{m+1,n}+\phi_{m,n}\over4}\right).
&(7.5\hbox{c})\cr}$$
The first index, $m$, on the fields represents the spatial lattice
site
and the second index, $n$ represents the temporal lattice site.
{\it Note the significant mnemonic for linear finite elements:
corresponding to directions in which derivatives are taken, a forward
difference
is taken, while in other directions, a forward average is taken.}
The index $m$ ranges from 0 to $M=L/\Delta$; all fields are taken to
be periodic
in the spatial lattice:
$$\phi_{0,n}=\phi_{M,n}.\eqno(7.6)$$
For technical reasons, which will be discussed below, we will take
$M$ to be
odd. As shown in Ref.~19, it is the space-averaged operators,
$$\eqalignno{\Phi_{m,n}&\equiv{1\over2}(\phi_{m,n}
+\phi_{m-1,n}),&\hbox{(7.7a)}
\cr
\Pi_{m,n}&\equiv{1\over2}(\pi_{m,n}+\pi_{m-1,n}),&\hbox{(7.7b)}\cr}$$
which obey the canonical equal-time commutation relations
$$\eqalign{[\Phi_{m,n},\Phi_{m',n}]&=[\Pi_{m,n},\Pi_{m',n}]=0,\cr
[\Phi_{m,n},\Pi_{m',n}]&={i\over\Delta}\delta_{m,m'}.\cr}\eqno(7.8)$$
These commutation relations are the discrete analogs of the continuum
equal-time commutation relations.

In principle, we can solve the system of difference equations (7.5);
for the purposes here, however, it is sufficient to expand in powers
of the
temporal lattice spacing $h$.  We do {\it not\/} expand in $\Delta$.
The expansions for $\phi_{m,n+1}$, $\pi_{m,n+1}$, and
$\Gamma_{m,n+1}$ are
$$\eqalign{\phi_{m,n+1}&=\phi_{m,n}+h A_m+h^2 B_m+O(h^3),\cr
\pi_{m,n+1}&=\pi_{m,n}+h C_m+h^2 D_m+O(h^3),\cr
\Gamma_{m,n+1}&=\Gamma_{m,n}+h E_m+h^2 F_m+O(h^3).\cr}\eqno(7.9)$$
Inserting (7.9) into (7.5) leads to a set of difference equations for
the
operators $A_m$, $B_m$, $C_m$, \dots. These equations all have the
same
generic form:
$$x_m+x_{m+1}=R_m,\eqno(7.10)$$
which, for periodic boundary conditions and $M$ odd,  has the general
solution
$$x_m={1\over2}\left[\sum_{k=m}^{M-1}(-1)^{k+m}R_k
-\sum_{k=0}^{m-1}(-1)^{k+m}
R_k\right]={1\over2}\sum_{k=m}^{M+m-1}(-1)^{k+m}R_k.\eqno(7.11)$$

Next, following the quantum-mechanical discussion in Sec.~II, we
introduce
a Fock-space representation for the canonical operators $\Phi_{m,n}$
and
$\Pi_{m,n}$ defined in (7.7):
$$\eqalignno{\Phi_{m,n}&=\sum_{k=1}^M\gamma_k(a_ke^{ikm
2\pi/m}+a_k^\dagger
e^{-ikm2\pi/M}),&\hbox{(7.12a)}\cr
\Pi_{m,n}&=\sum_{k=1}^M{i\over2\gamma_k L}(-a_ke^{ikm
2\pi/m}+a_k^\dagger
e^{-ikm2\pi/M}),&\hbox{(7.12b)}\cr}$$
where
$$\eqalignno{[a_k,a_l^\dagger]&=\delta_{k,l},&\hbox{(7.13a)}\cr
[a_k,a_l]&=[a_k^\dagger,a_l^\dagger]=0.&\hbox{(7.13b)}\cr}$$
In (7.12) $\gamma_k$, $k=1,2,\dots, M$, are arbitrary parameters
which later will be fixed by a variational argument, similar to that
given in Sec.~II\null.
Recall that the spatial size of the lattice is $L=M\Delta$.

As a consequence of (7.13), the equal-time commutation relations
(7.8) are satisfied at the initial time $n$ provided that
$$\gamma_k^2=\gamma_{M-k}^2.\eqno(7.14).$$
It is crucial that as a consequence of the operator difference
equations (7.5) the equal time commutation relations (7.8) are
preserved
at all subsequent lattice sites, in particular those at $n+1$.

Using (7.12) we find easily
$${1\over2}(\Gamma_{m-1,n}+\Gamma_{m,n})={2i\over\Delta}
\sum_{k=1}^M\gamma_k\tan{k\pi\over M}(a_ke^{ikm 2\pi/m}-a_k^\dagger
e^{-ikm2\pi/M}),\eqno(7.15)$$
It is easy to verify the expected equal-time commutation relation
between
${1\over2}(\Gamma_{m-1,n}+\Gamma_{m,n})$ and $\Phi_{m,n}$:
$$[\Phi_{m,n},{1\over2}(\Gamma_{m-1,n}+\Gamma_{m,n})]=0.\eqno(7.16)$$
We will merely display the order-$h$ coefficients:
$$\eqalignno{A_m&=\Pi_m,&\hbox{(7.17a)}\cr
{1\over2}(E_{m-1}+E_{m})&={1\over\Delta L}
\sum_{k=1}^M{1\over\gamma_k}\tan{k\pi\over M}(a_ke^{ikm
2\pi/m}+a_k^\dagger
e^{-ikm2\pi/M}),&\hbox{(7.17b)}\cr
{1\over2}(C_{m-1}+C_{m})&=-{4\over\Delta^2}
\sum_{k=1}^M\gamma_k\tan^2{k\pi\over M}(a_ke^{ikm 2\pi/m}+a_k^\dagger
e^{-ikm2\pi/M})\cr&\qquad+f(\Phi_{m,n}).&\hbox{(7.17c)}\cr}$$
We do not bother to display the solutions for $B_m$, $D_m$, and
$F_m$;
however, they can be calculated similarly.

Consider first a free theory, for which $f=-\mu^2\phi$.  We extract
spectral
information by taking matrix elements of (7.9) between the Fock
vacuum and a one-particle state with lattice momentum $l$:
$$\langle 1, l|=\langle 0|a_l.\eqno(7.18)$$
We compare these matrix elements with an assumed approximate
exponential
time dependence with a single frequency:
$$\eqalign{\langle 1,l|\Phi_{m,n+1}|0\rangle&\approx e^{i\omega_l h}
\langle 1,l|\Phi_{m,n}|0\rangle,\cr
\langle 1,l|\Pi_{m,n+1}|0\rangle&\approx e^{i\omega_l h}
\langle 1,l|\Pi_{m,n}|0\rangle.\cr}\eqno(7.19)$$
The $O(h^0)$ and $O(h^1)$ equations give a relation between the
frequency
$\omega_l$ and the variational parameter $\gamma_l$,
$$\omega_l={1\over2 L\gamma_l},\eqno(7.20)$$
and the dispersion relation
$$\omega_l^2=\mu^2+{4\over\Delta^2}\tan^2{\pi l\over M}.\eqno(7.21)$$
In the continuum limit, $\Delta\to0$, $M\to\infty$, and
$L=M\Delta\to\infty$,
the lattice equivalent of the continuum momentum $p$ is
$$p={2\pi l\over M\Delta}.\eqno(7.22)$$
Thus, we recover the continuum dispersion relation
$$\omega^2=\mu^2+p^2.\eqno(7.23)$$

It is remarkable that if we include the two additional equations
coming
from the $O(h^2)$ terms in $\Phi_{m,n+1}$ and $\Pi_{m,n+1}$ as well
as the three
further equations coming from $\Gamma_{m,n+1}$ only redundant
information
is supplied.

Consider now an interacting theory for which
$$V={\mu^2\phi^2\over2}+{g\over2N}\phi^{2N}.\eqno(7.24)$$
Following the procedure described above, we find that (7.20) still
holds, but the dispersion relation (7.21) is replaced by
$$\omega_l^2=m_{\rm ren}^2+{4\over\Delta^2}\tan^2{\pi l\over
M},\eqno(7.25)$$
where $$ m_{\rm ren}^2=\mu^2+(2N-1)!!\,g X^{N-1},
\quad X=\sum_{k=1}^M\gamma_k^2,\eqno(7.26)$$
In the continuum limit, an asymptotic analysis$^{20}$ serves to
evaluate $X$:
$$X={1\over2\pi}\ln\left({4\over m_{\rm
ren}\Delta}\right).\eqno(7.27)$$
An equation for the renormalized mass is thus obtained when we
substitute
(7.27) into (7.26):
$$m_{\rm ren}^2=\mu^2+{(2N-1)!!\,g\over(2\pi)^{N-1}}\left[\ln\left(
{4\over m_{\rm ren}\Delta}\right)\right]^{N-1}.\eqno(7.28)$$
This nonperturbative result closely resembles the formula in
continuum
perturbation theory:
$$m_{\rm ren}^2=\mu^2+{(2N-1)!!\,g\over(2\pi)^{N-1}}\left[\ln\left(
{\Lambda\over \mu}\right)\right]^{N-1}.\eqno(7.29)$$
The correspondence is provided by the identification of the momentum
cutoff
$\Lambda$ with $\pi/\Delta$.

We close this section by extending this calculation to the
sine-Gordon
model, for which
$$V={\mu^2\over g^2}\cos g\phi.\eqno(7.30)$$
The corresponding force is
$$f=-{\mu^2\over g}\sin
g\phi=-\mu^2\sum_{N=0}^\infty{(-1)^Ng^{2N}\phi^{2N+1}
\over(2N+1)!}.\eqno(7.31)$$
Each term in the sum in (7.31) gives a contribution to the
renormalized mass
of the form given in (7.26), so the formula for the renormalized mass
is
therefore
$$m_{\rm ren}^2=\mu^2\sum_{N=0}^\infty{(-1)^Ng^{2N}(2N+1)!!\,
X^N\over
(2N+1)!}=\mu^2 e^{-g^2 X/2},\eqno(7.32)$$
where $X$ is given in (7.26).  The dispersion relation is
$$\omega^2_l={4\over\Delta^2}\tan^2{\pi l\over M}+\mu^2 e^{-g^2
X/2}.\eqno(7.33) $$
An asymptotic analysis in the continuum limit leads once again to
(7.27),
so a simple calculation yields the following relation between the
renormalized and the unrenormalized masses:
$$m_{\rm ren}=\mu
\left({\mu\Delta\over4}\right)^{g^2/(8\pi-g^2)}.\eqno(7.34)$$
This is the characteristic power-law renormalization found in the
conventional
treatments of the sine-Gordon model$^{21}$,
and reduces to the perturbative result
of Coleman$^{22}$ when $g^2/8\pi$ is small.
\vfill\eject
\noindent{\bf VIII.~THE DIRAC EQUATION AND FERMION DOUBLING}
\bigskip

The finite-element lattice Dirac equation in $3+1$ dimensions is
$${i\gamma^0\over h}(\psi_{\overline{\bf m},n+1} -\psi_{\overline{\bf
m},
n})+{i\gamma^j\over\Delta}(\psi_{m_j+1, \overline{\bf
m}_\perp,\overline{n}}
-\psi_{m_j,\overline{\bf m}_\perp,\overline{n}})
+\mu\psi_{\overline{\bf m},
\overline{n}}=0.\eqno(8.1)$$
Here the overbar represents a forward average over that coordinate:
$$x_{\overline m}={1\over2}(x_{m+1}+x_m),$$
and the notation ${\bf\overline m_\perp}$ means that all spatial
coordinates
but $m_j$ are averaged. Let us begin by finding the momentum-space
spinors,
the eigenvectors of the transfer matrix.  That is, write for a plane
wave
at time $n$
$$\psi_{\overline{\bf m},n}=u_n e^{-i{\bf p}
\cdot{\bf m} 2\pi/M},\eqno(8.2\hbox{a})$$
and at time $n+1$
$$\psi_{\overline{\bf m},n+1}=
u_{n+1} e^{-i{\bf p}\cdot{\bf m} 2\pi/M}.\eqno(8.2\hbox{b})$$
The transfer matrix $T$ is defined by
$$ u_{n+1}=T u_n.\eqno(8.3) $$
By substituting (8.2a) and (8.2b) into the Dirac equation (8.1)
we easily find that
$$\eqalignno{
T&=\left({i\gamma^0\over h}+ {\vec\gamma
\cdot{\bf t}\over\Delta}+{\mu\over2}\right)^{-1}
\left({i\gamma^0\over h}-{\vec\gamma
\cdot{\bf t}\over\Delta}-{\mu\over2}\right)\cr
&=\left(1+{\mu^2h^2\over4}+{h^2\over\Delta^2}t^2\right)^{-1}
\left(1-{\mu^2h^2\over4}
-{h^2\over\Delta^2}t^2
+{2h\over\Delta}i\gamma^0\vec\gamma\cdot{\bf t}
+\mu hi\gamma^0\right),&(8.4)}$$
where
$$ {\bf t}={\bf t}_{\bf p},\quad ({\bf t}_{\bf p})_i=\tan p_i
\pi/M.\eqno(8.5)$$
Let us adopt a representation of the Dirac matrices in which
$$ \gamma^0=\left(
\matrix{
1&0\cr
0&-1\cr}\right),\quad
i\gamma^0\gamma^j=\sigma^{0j}=i\left(
\matrix{
0&\sigma^j\cr
\sigma^j&0\cr}\right).\eqno(8.6)$$
Then the eigenvalues of $T$ are easily found:
$$ \lambda={1\pm i h\tilde\omega/2\over1\mp i h
\tilde\omega/2}.\eqno(8.7)$$
Here $\tilde\omega$ is an abbreviation for
$$ \tilde\omega=\sqrt{{4t^2\over\Delta^2}+\mu^2},\eqno(8.8) $$
which is exactly the same as the dispersion relation (7.21).
It is obvious that $\lambda$ has modulus unity, so it can
be written in the form $\lambda=\exp(\pm i\omega h)$,
where $\omega$ is, of course,
a function of $h$. (The relation between $\omega$ and $\tilde\omega$
is
$\tilde\omega={2\over h}\tan{h\omega\over2}$.)
The corresponding eigenvectors may also be
found straightforwardly.  They are to be normalized according to
$$ u_\pm^\dagger\gamma^0 u_{\pm}=\pm1.\eqno(8.9)$$
They are
$$ u_\pm=\left(
\matrix{
\pm{[(\tilde\omega\pm \mu)/2\mu]}^{1/2}
\vec\sigma\cdot{\bf t}/t\cr
{[(\tilde\omega\mp \mu)/2\mu]}^{1/2}
\cr}
\right) \chi,\eqno(8.10) $$
where $\chi$ is a two-component, rest-frame spinor, normalized by
$\chi^\dagger \chi=1$.
Thus, with
$$ i\gamma_5=\left(\matrix{
0&&1\cr
1&&0\cr}\right),\eqno(8.11)$$
we have
$$ u_\pm=\left[\left({\tilde\omega+\mu\over2\mu}\right)^{1/2}
\pm i\gamma_5{\vec\sigma
\cdot{\bf t}\over t}\left({\tilde\omega-\mu\over 2\mu}
\right)^{1/2}\right]u^{(0)}_\pm,\eqno(8.12)$$
where $u^{(0)}_\pm$ is a four-component rest-frame spinor with
$\gamma^0$ eigenvalue of $\pm1$.  Therefore, in terms of the spinors
$\tilde u_+({\bf p})=u_+({\bf p})$,
$\tilde u_-({\bf p})=u_-(-{\bf p})$,
we have the completeness relations
$$ \sum_{\rm{spins}}\tilde u_\pm \tilde u^\dagger_\pm
\gamma^0=\pm{1\over2\mu}\left(
\mu\pm\gamma^0\tilde\omega\mp{2\vec\gamma
\cdot{\bf t}\over \Delta}\right), \eqno(8.13)$$
which in the continuum limit reduces to $\pm(\mu\mp\gamma\cdot
p)/2\mu$.
We have the same result on the lattice, provided we define
$$ \tilde p^0=\tilde\omega,\quad \tilde{\bf p}=
{2{\bf t}\over\Delta}.\eqno(8.14)$$

All of this tells us that the momentum expansion of the Dirac field
has the form
$$\psi_{\overline{\bf m},n}=\sum_{s,{\bf p}}\sqrt{{\mu\over\tilde
\omega}}
\left(b_{{\bf p},s}u_{{\bf p},s}e^{i{\bf p}\cdot{\bf m} 2\pi/M}
+d^\dagger_{{\bf p},s}v_{{\bf p},s}
e^{-i{\bf p}\cdot{\bf m} 2\pi/M}\right), \eqno(8.15)$$
where we now use the standard notation
$u=i\gamma_5\tilde u_-$, $v= i\gamma_5\tilde u_+$,
with the usual interpretation that $d^\dagger$ creates a positive
energy
positron, while $b$ annihilates a positive energy electron.
The canonical lattice anticommutation relations
$$ \{\psi_{\overline{\bf m},n}, \psi^\dagger_{\overline{\bf m}',n}\}
={1\over\Delta^3}\delta_{{\bf m},{\bf m}'}\eqno(8.16)$$
will now be satisfied if
$$ \{b_{{\bf p},s},b^\dagger_{{\bf p}',s'}\}
={1\over L^3}\delta_{{\bf p},
{\bf p}'}\delta_{s,s'},\quad
\{d_{{\bf p},s},d^\dagger_{{\bf p}',s'}\}
={1\over L^3}\delta_{{\bf p},
{\bf p}'}\delta_{s,s'},\eqno(8.17)$$
and all other anticommutators of these operators vanish.

The unitarity of $T$ is sufficient to establish the
unitarity of the fermion sector in the noninteracting theory.
For further details see Refs.~23 and 24.

It is apparent that the dispersion relation (8.8) solves the
fermion doubling problem; that is, $\omega^2$ assumes the value
$\mu^2$ at $\bf p=0$ (mod $M$) and nowhere else.$^{25}$ In this
discussion
we have assumed, as in Sec.~VII, that $M$ is odd so that the Dirac
field is periodic; however, the same conclusion would hold if
antiperiodic boundary conditions were used. In the dispersion
relation
one would simply replace $p$ by $p+{1\over2}$.

Let us conclude this section by summarizing the properties of the
linear finite-element Dirac equation (8.1):
\item{1.} It is unitary in that the equal-time anticommutation
relations
are exactly preserved in time.
\item{2.} It may be derived from an Hermitian action.
\item{3.} There is no fermion doubling.
\item{4.} The difference equation is local, in that only
nearest-neighbor
terms appear.
\item{5.} It is chirally symmetric in the massless limit.

\noindent For a complete discussion of these points see Ref.~23.

The no-go theorems of Karsten and Smit, Nielsen and Ninomiya, and
Rabin$^{26}$
are avoided because the time development operator is nonlocal, which
arises because undifferentiated fields appear as averages.
\vfill\eject
\noindent{\bf IX.~DIRAC EQUATION WITH INTERACTIONS AND THE SCHWINGER
MODEL}
\bigskip

In the last two sections of this review, we will discuss interactions
of fermions with gauge fields.  In this section we will concentrate
on
the Dirac equation, interacting with either Abelian or nonabelian
gauge fields, and make application to electrodynamics, in particular
to
the Schwinger model.  In the next section the nonabelian interactions
of the gauge fields among themselves will be derived.
(The original treatment of finite-element electrodynamics was given
in Ref.~27.)
For simplicity, initially our discussion will be restricted to (1+1)
dimensions.
\bigskip
\noindent{\bf A.~Equations of Motion in the Continuum}
\medskip

We begin by recalling the form of the continuum field equations
of a nonabelian gauge field $A_\mu$ coupled to a fermion field
$\psi$.
Let us start with the free Dirac equation
\def\sla{\raise.15ex\hbox{$/$}\kern-.57em}
$$(i\sla\partial+\mu)\psi=0.\eqno(9.1)$$
Equation (9.1) is invariant under an infinitesimal gauge
transformation
$$\psi\to\psi+\delta\psi,\quad \delta\psi=i g \delta\omega\psi,\quad
\delta\omega=\delta\omega_a T_a,\eqno(9.2)$$
provided $\delta\omega$ is constant.  Here $T$ is the generator
of the gauge group.  If $\delta\omega$
is not constant, we can restore the invariance  by adding an
interaction
term to the Dirac equation,
\def\slsh{\raise.15ex\hbox{$/$}\kern-.7em}
$$(i\sla\partial+g\slsh A+\mu)\psi=0,\quad A^\mu=A^\mu_a
T_a,\eqno(9.3)$$
provided $A$ transforms according to
$$A_\mu\to A_\mu+\delta A_\mu, \quad \delta A_\mu =
\partial_\mu\delta\omega +ig[\delta\omega, A_\mu].\eqno(9.4)$$
Under (9.4) the curl of $A$ is not covariant, so we must add a
suitable
interaction term to construct the field strength:
$$F_{\mu\nu}=\partial_\mu A_\nu-\partial_\nu A_\mu-ig[A_\mu,
A_\nu].\eqno(9.5)$$
The field strength transforms covariantly:
$$F_{\mu\nu}\to F_{\mu\nu}+\delta F_{\mu\nu},\quad \delta F_{\mu\nu}
=ig[\delta\omega, F_{\mu\nu}].\eqno(9.6)$$
Finally, because the current $$j^\mu=g\bar\psi T\gamma^\mu
\psi\eqno(9.7)$$
transforms covariantly, $$j^\mu\to j^\mu+\delta j^\mu,\quad \delta
j^\mu
= ig [\delta\omega, j^\mu],\eqno(9.8)$$
we must include the interaction term in the Yang-Mills equation,
$$D_\nu F^{\mu\nu}=j^\mu,\qquad\hbox{where}\qquad D_\nu=\partial_\nu
-ig[A_\nu,\quad].\eqno(9.9)$$
\bigskip
\noindent{\bf B.~Free Lattice Equations of Motion}
\medskip
\def\th{\theta}
\def\de{\delta}
\def\om{\omega}
\def\ga{\gamma}

In $1+1$ dimensions,  the free Dirac equation (8.1) reads
$${i\gamma^0\over h}(\phi_{m,n+1}-\phi_{m,n})
+{i\gamma^1\over \Delta}(\theta_{m+1,n}-\theta_{m,n})
+{\mu\over2}(\theta_{m+1,n}+\theta_{m,n})=0.\eqno(9.10)$$
Here we have used the abbreviations
$$\eqalignno{\phi_{m,n}&={1\over2}(\psi_{m+1,n} +\psi_{m,n}),
&(9.11\hbox{a})\cr
\theta_{m,n}&={1\over2}(\psi_{m,n+1}+\psi_{m,n}).
&(9.11\hbox{b})\cr}$$

Similarly, the field strength $E=F_{01}$ is constructed as
$$\tilde E^{(0)}_{m,n}={1\over h}(C_{m+1,n+1}-C_{m+1,n})
-{1\over \Delta}(B_{m+1,n+1}-B_{m,n+1}),\eqno(9.12)$$
where
$$\eqalignno{B_{m,n}&={1\over2}[(A_0)_{m,n}
+(A_0)_{m,n-1}],&(9.13\hbox{a})\cr
C_{m,n}&={1\over2}[(A_1)_{m,n}+(A_1)_{m-1,n}].&(9.13\hbox{b})\cr}$$
In (9.12) the tilde signifies the average over the four adjacent
lattice sites:
$$\tilde E^{(0)}_{m,n}={1\over4}(E^{(0)}_{m+1,n+1}+E^{(0)}_{m+1,n}
+E^{(0)}_{m,n+1}+E^{(0)}_{m,n})\equiv E^{(0)}_{\overline{m},\overline
n},
\eqno(9.14)$$
and the $(0)$ superscript is a reminder that this is the free field
strength.

Finally, the free Yang-Mills equations driven by a current are
$$\eqalignno{{1\over h}(F_{m+1,n}-F_{m,n})&=-\tilde \jmath^0_{m,n},
&(9.15\hbox{a})\cr
{1\over h}(G_{m,n+1}-G_{m,n})&=\tilde
\jmath^1_{m,n},&(9.15\hbox{b})\cr}$$
where
$$\eqalignno{F_{m,n}&={1\over2}(E_{m,n+1}+E_{m,n}),&(9.16\hbox{a})\cr
G_{m,n}&={1\over2}(E_{m+1,n}+E_{m,n}),&(9.16\hbox{b})\cr}$$
and $$\tilde \jmath^\mu_{m,n}=g\bar\Psi_{m,n} T\gamma^\mu\Psi_{m,n}.
\eqno(9.17)$$
Here we have used
$\Psi=\tilde\psi$. (We have, of course, anticipated the interaction
with the fermion by including $j^\mu$.)
\bigskip
\noindent{\bf C.~Constructing the Interaction Term Involving $A^1$}
\medskip

We now proceed to construct the interaction terms for the Dirac
equation (9.10).
We begin by recognizing that the gauge transformation of the fermion
on the lattice is
$$\delta\Psi_{m,n}=ig\delta\omega_{m,n}\Psi_{m,n},\eqno(9.18)$$
which is the appropriate lattice version of (9.2) because (9.18)
guarantees
that the mass term in (9.10) transforms covariantly, that is, by the
same rule.  (We also note that then the current in (9.17) transforms
covariantly.)  We can regard (9.18) as a difference equation for
$\delta\psi_{m,n}$; in particular, we find
$$\delta\theta_{m,n}=ig\delta\omega_{m,n}\th_{m,n}
+{ig\over2}\sum_{m'=m+1}^{M+m}
(-1)^{m+m'}(\delta\omega_{m',n}-\delta\omega_{m'-1,n})\theta_{m',n}.
\eqno(9.19)$$
Here, we have used either periodic or antiperiodic boundary
conditions
depending on the size of the lattice:
$$\psi_{m+M,n}=(-1)^{M+1}\psi_{m,n},\quad
\delta\psi_{m+M,n}=(-1)^{M+1}\delta\psi_{m,n},\eqno(9.20)$$
where $M$ is the number of spatial lattice sites.  The boson
variables such
as $\de\om_{m,n}$ will be assumed to be periodic.

It is evident that the second term in (9.10) does not transform
covariantly
under (9.19); as in the continuum, an interaction term must be added.
Now the free field strength in (9.12) is {\it invariant\/} under
the transformations
$$\eqalignno{(\delta A_0)_{m,n}&={1\over 2h}(\delta\Lambda_{m+1,n+1}
+\delta\Lambda_{m,n+1}-\delta\Lambda_{m+1,n}
-\delta\Lambda_{m,n}),&(9.21
\hbox{a})\cr
(\delta A_1)_{m,n}&={1\over
2\Delta}(\delta\Lambda_{m+1,n+1}+\delta\Lambda_{m+1,
n}-\delta\Lambda_{m,n+1}-\delta\Lambda_{m,n}),&(9.21\hbox{b})\cr}$$
The finite-element connection between $\delta\Lambda_{m,n}$ and
$\delta\omega_{m,n}$ is $$\delta\omega_{m,n}=
{1\over4}(\delta\Lambda_{m+1,n+1}
+\delta\Lambda_{m+1,n}+\delta\Lambda_{m,n+1}
+\delta\Lambda_{m+1,n}).\eqno(9.22)$$
Then the scalar and vector potentials, (9.13a) and (9.13b), transform
by
$$\eqalignno{\delta^{(0)}B_{m,n}&={1\over
h}(\delta\om_{m,n}-\de\om_{m,n-1}),
&(9.23\hbox{a})\cr
\de^{(0)}C_{m,n}&={1\over \Delta}
(\de\om_{m,n}-\de\om_{m-1,n}).&(9.23\hbox{b})}$$
Here the superscript (0) reminds us that further transformations of
$B$ and
$C$ will have to be deduced.

In the rest of this subsection we will examine that portion of the
Dirac
equation proportional to $\gamma^1$.  A short calculation reveals
that
under (9.19)
$$\eqalignno{\de\left[{i\ga^1\over
\Delta}(\th_{m+1,n}-\th_{m,n})\right]&=
{i\ga^1\over
\Delta}ig{\bigg[}\delta\om_{m,n}(\th_{m+1,n}-\th_{m,n})\cr
&\quad-\sum_{m'=m+1}^{m+M}(-1)^{m+m'}(\de\om_{m',n}
-\de\om_{m'-1,n})\th_{m',n}{\bigg]}.&(9.24)}$$
The first term here expresses the desired covariance of this term in
the
Dirac equation.  The second term will be cancelled if we introduce
an interaction term
$$I^{(1)}_{m,n}={i\ga^1\over \Delta}ig\Delta
\sum_{m'=m+1}^{m+M}(-1)^{m+m'}C_{m',n}
\theta_{m',n},\eqno(9.25)$$
and vary it with respect to $C$ according to (9.23b).  Indeed, in the
continuum limit, this term reduces to the corresponding interaction
term in (9.3) because
$$\sum_{m'=m+1}^{m+M}(-1)^{m+m'}g_{m'}\to
-g(m\Delta)\quad\hbox{as}\quad \Delta\to0,\quad m\to\infty,\quad
M\to\infty,
\eqno(2.26)$$
if $g_{M+m}=(-1)^{M+1}g_m$. But, on the lattice, we are not finished,
for
we have neither yet varied (9.25) with respect to $\th$ according to
(9.19),
nor achieved the covariance of the interaction term.

A straightforward calculation now reveals that
$$\delta_\theta
I^{(1)}_{m,n}=ig\de\om_{m,n}I^{(1)}_{m,n}-\delta^{(1)}I_{
m,n}^{(1)} -\delta^{(0)}I_{m,n}^{(2)}.\eqno(9.27)$$
Here the first term is the required
covariance term, the second term involves a new variation of $C$,
$$\de^{(1)}C_{m,n}=ig{1\over2}[\de\om_{m,n}+\de\om_{m-1,n},C_{m,n}],
\eqno(9.28)$$
and the third term is the $\delta^{(0)}$ variation of a new
interaction term
$$I^{(2)}_{m,n}=-{i\ga^1\over 2\Delta}(ig\Delta)^2
\sum_{m'=m+1}^{m+M}\sum_{m''=m'+1}
^{m'+M-1}(-1)^{m+m''}C_{m',n}C_{m'',n}\th_{m'',n}.\eqno(9.29)$$
It is easy to see that (9.29) vanishes in the continuum limit, while
the
variation (9.28) reduces to the second term in (9.4) in that same
limit.
Once again, we are not finished: we still need to vary $I^{(2)}$ with
respect to $\de\th$ and $\de^{(1)}C$, and produce the required
covariance
of $I^{(2)}$.

Clearly, this process of adding successive interaction terms and
evermore
$C$ variations never terminates. But it is easy to discern the
general
pattern.  The easiest way to express $I^{(N)}$, where $N$ is the
order
in $g\Delta C$, is by the following inductive formula: (here and in
the rest
of this section we delete the time index $n$, since all variables are
evaluated at that time)
$$N\ge1:\quad I^{(N)}_m={1\over2}\sum_{m'=m+1}^{m+M}(-1)^{m+m'}
\sum_{k=1}^{N}{1\over k!}(-ig\Delta C_{m'})^k
I_{m'}^{(N-k)},\eqno(9.30)$$
where we define $I^{(0)}_m=-2i\ga^1\th_m/\Delta$.
The gauge transformations are given by (9.19) and
$$\eqalignno{k\ne1:\quad
\de^{(k)}C_m&={(ig\Delta)^k\over \Delta}{{\cal B}_k\over k!}
\underbrace{[\dots[\de\om_m-
\de\om_{m-1},C_m],\dots,C_m]}_{k {\rm\;nested\; commutators}}
&(9.31\hbox{a})\cr
\de^{(1)}C_m&={ig\over2}[\de\om_m+\de\om_{m-1},C_m],
&(9.31\hbox{b})\cr}$$
where ${\cal B}_k$ is the $k$th Bernoulli number.
The required covariance statement
$$\de_\th I^{(N)}_m+\sum_{k=0}^N\de^{(k)}I^{(N-k+1)}_m=ig\de\om_m
I^{(N)}_m
\eqno(9.32)$$
is proved in Ref.~28.

{}From (9.30) we can derive an ``integral equation'' satisfied by
the full interaction term for the vector potential,
$I=\sum_{N=1}^\infty I^{(N)}$:
$$I_m={1\over2}\sum_{m'=m+1}^{m+M}(-1)^{m+m'}\left(e^{-ig\Delta
C_{m'}}-1\right) \left(I_{m'}^{(0)}+I_{m'}\right).\eqno(9.33)$$
{}From this, a difference equation can be immediately derived:
$$I_m+e^{ig\Delta C_m}I_{m-1}={2i\ga^1\over
\Delta}\left(1-e^{ig\Delta C_m}\right)\th_m.\eqno(9.34)$$
\bigskip
\noindent{\bf D.~Scalar Potential Part of Dirac Equation}
\medskip

The procedure is now clear.  We start from the part of the free Dirac
equation
(9.10) proportional to $\gamma^0$,
$${i\ga^0\over h}(\phi_{n+1}-\phi_n).\eqno(9.35)$$
Here we have dropped the spatial index $m$ because throughout this
section
all variables will be evaluated at the same spatial coordinate.  Now,
we need
to solve the fermion transformation equation (9.18) for $\delta\phi$.
We immediately find
$$\eqalignno{\de\phi_n&=ig\de\om_n\phi_n+(-1)^n
(\de\phi_0-ig\de\om_0\phi_0)
\cr&\quad-ig\sum_{n'=1}^n(-1)^{n+n'}(\de\om_{n'}
-\de\om_{n'-1})\phi_{n'}.&(9.36)}$$
At this point we make a slight variation on the procedure of Ref.~27.
We will choose as an initial condition on the $\phi$ variation
$$\de\phi_0={ig\over2}(\de\om_0+\de\om_{-1})\phi_0,\eqno(9.37)$$
which will simplify the form of subsequent formulas.  The reason for
the choice (9.37) (or the slightly different choice made in Ref.~27)
is to ensure that the lattice variation of the time difference of
$\phi$ and hence the Dirac equation have the correct continuum limit.
The latter is obtained from
$$2\sum_{n'=1}^n(-1)^{n+n'}f_{n'}+(-1)^{n}
f_0\to f(nh),\quad\hbox{ as }\quad h\to0,\quad
n\to\infty.\eqno(9.38)$$
Using (9.37) we write (9.36) as
$$\de\phi_n=ig\de\om_n\phi_n-ig\mathop{{\sum}'}_{n=0}^{n}
(-1)^{n+n'}(\de\om_{n'} -\de\om_{n'-1})\phi_{n'},\eqno(9.39)$$
where the prime on the sum signifies that the 0th term is counted
with half
weight:
$$\mathop{{\sum}'}_{n'=0}^n
f_{n'}=\sum_{n=1}^nf_{n'}+{1\over2}f_0.\eqno(9.40)$$
The variation of (9.35) is immediate:
$$\eqalignno{\de\left[{i\ga^0\over h}(\phi_{n+1}-\phi_n)\right]
&={i\ga^0\over h}
ig\de\om_n(\phi_{n+1}-\phi_n)\cr
&\quad+2ig{i\ga^0\over
h}\mathop{{\sum}'}_{n'=0}^n(-1)^{n+n'}(\de\om_{n'}
-\de\om_{n'-1})\phi_{n'}.&(9.41)}$$
The first term here is the required
covariance of (9.35), while the second term on the right-hand side of
(9.41) is cancelled by the variation of the following interaction
term,
$$K^{(1)}_n=-2(igh){i\ga^0\over
h}\mathop{{\sum}'}_{n'=0}^n(-1)^{n+n'}
B_{n'}\phi_{n'},\eqno(9.42)$$
under (9.23a),
$$\delta^{(0)}B_n={1\over h}(\de\om_n-\de\om_{n-1}).\eqno(9.43)$$
Of course, using (9.38), we find that $K^{(1)}$ reduces to the
appropriate
interaction term in the continuum Dirac equation.

However, here again we are not finished.  We must vary $K^{(1)}$ with
respect to
$\de\th$.  Doing so necessitates the introduction of a new field
variation,
$$\eqalignno{n\ge1:\quad
\de^{(1)}B_n&={ig\over2}[\de\om_n+\de\om_{n-1},B_n],
&(9.44\hbox{a})\cr
n=0:\quad \de^{(1)}B_0&={ig\over2}[\de\om_0+\de\om_{-1},B_0]
+{ig\over4}[\de\om_0-\de\om_{-1},B_0],&(9.44\hbox{b})\cr}$$
which reduces to the appropriate portion of the
continuum variation (9.4) in the continuum limit,
and a new interaction term, $$K^{(2)}_n=-{i\gamma^0\over h}2(igh)^2
\mathop{{\sum}'}_{n'=0}^n\mathop{{\sum}''}_{n''=0}^{n'}
(-1)^{n+n''}B_{n'}B_{n''}\phi_{n''},\eqno(9.45)$$
which vanishes in the continuum limit.  Here
$$\eqalignno{n'\ge1:\quad \mathop{{\sum}''}_{n''=0}^{n'}
f_{n''}&={1\over2}f_0
+\sum_{n''=1}^{n'-1}f_{n''}+{1\over2}f_{n'},\cr
\mathop{{\sum}''}_{n''=0}^0 f_{n''}&=0.&(9.46)\cr}$$

Once again, this iterative procedure continues indefinitely.
We may again write the general result in terms of an inductive
formula.  The order $N$ interaction term is
$$\eqalignno{K^{(N)}_n&=-\sum_{n'=1}^n(-1)^{n+n'}\sum_{k=1}^N
{(-igh)^k\over k!}B^k_{n'}K^{(N-k)}_{n'}\cr
&\quad+{1\over h}(igh)^N{1\over
2^{N}N!}(-1)^nB_0^NK^{(0)}_0,&(9.47)\cr }$$
where $K^{(0)}_n=-2i\ga^0\phi_n/h$.
In Ref.~28 we show that these transform according to the required
law,
$$\de_\phi K^{(N)}_n+\sum_{k=0}^N\de^{(k)}K^{(N-k+1)}_n=ig\de\om_n
K_n^{(N)
},\eqno(9.48)$$
where $\de\phi_n$ is given by (9.39) and
$$\eqalignno{k\ne1:\quad
\de^{(k)}B_n&={(igh)^k\over h}{{\cal B}_k\over
k!}\underbrace{[\dots[\de\om_n
-\de\om_{n-1},B_n],\dots,B_n]}_{k\rm{\;nested\; commutators}},
\quad n\ne0,&(9.49\hbox{a})\cr
\de^{(1)}B_n&={ig\over2}[\de\om_n+\de\om_{n-1},B_n],\quad
n\ne0,&(9.49
\hbox{b})\cr
k\ne 1:\quad \de^{(k)}B_0&={(igh)^k\over h}(-1)^k{{\cal B}_k\over k!
2^k}
\underbrace{[\dots[\de\om_0-\de\om_{-1},B_0],\dots,B_0]}_{k{\rm\;
nested\;
commutators}},&(9.49\hbox{c})\cr
\de^{(1)}B_0&={ig\over2}[\de\om_0+\de\om_{-1},B_0]-ig{{\cal
B}_1\over2}
[\de\om_0-\de\om_{-1},B_0],&(9.49\hbox{d})\cr}$$
where, again, ${\cal B}_k$ is the $k$th Bernoulli number.

As before, from (9.47) an ``integral equation'' can be immediately
obtained for the full interaction term with the scalar potential,
$K_n=\sum_{N=1}^\infty K_n^{(N)}$:
$$\eqalignno{K_n&=-\sum_{n'=1}^n(-1)^{n+n'}
\left(e^{-ighB_{n'}}-1\right)
\left(K_{n'}+K_{n'}^{(0)}\right)\cr
&\quad+(-1)^n\left(e^{ighB_0/2}-1\right)K_0^{(0)},&(9.50)\cr}$$
which is equivalent to the difference equation
$$K_n+e^{ighB_n}K_{n-1}={2i\ga^0\over
h}\left(1-e^{ighB_n}\right)\phi_n.\eqno(9.51)$$

The full lattice Dirac equation is given by (9.10), (9.30), and
(9.47):
$${i\ga^0\over h}(\phi_{m,n+1}-\phi_{m,n})+{i\ga^1\over
\Delta}(\th_{m+1,n} -\th_{m,n})+{\mu\over2}(\phi_{m,n+1}+\phi_{m,n})+
(I_{m,n}+K_{m,n})=0.\eqno(9.52)$$
Note that (9.52) gives $\psi_{m,n+1}$ in terms of fields at time
$n$ and earlier, so that this difference equation may be solved by
time stepping through the lattice.
\bigskip
\noindent{\bf E.~Unitarity of the Dirac Equation in the Temporal
Gauge}
\medskip

For the case of Abelian electrodynamics, the Dirac equation may be
written
explicitly. (For a complete discussion, see Ref.~27.)
The generalization to four dimensions is
immediate, and the result may be expressed in terms
of the spatially averaged electron field as follows:$^{24}$
$$\eqalignno{
{i\over h}
&(\psi_{\overline{\bf m},n+1}-\psi_{\overline{\bf m},n})
-{2i\gamma^0\gamma^j\over\Delta}\left[ \sum_{m'_j=m_j+1}^M
(-1)^{m_j+m_j'}
\psi_{\overline{m}_j',\overline{\bf m}_\perp,\overline{n}}
-\sum_{m'_j=1}^{m_j-1} (-1)^{m_j+m_j'}
\psi_{\overline{m}_j',\overline{\bf m}_\perp,\overline{n}}
\right]\cr
&+{\mu\gamma^0\over2}
(\psi_{\overline{\bf m},n+1}+
\psi_{\overline{\bf m},n})+2{\gamma^0\gamma^j\over\Delta}
\sum_{m_j'}\alpha^{(j)}_{{\bf m}_\perp;m_j,m_j'}\psi_{
\overline{m}_j',\overline{\bf m}_\perp,\overline{n}}=0,&(9.53)\cr}$$
where a sum over the repeated index $j$ is understood.
Here, we have adopted a temporal gauge, $A^0=0$, and expressed the
interaction
in terms of (only the $j$th index is explicit and the
spatial coordinates refer to the $j$th direction)
$$\eqalignno{
\alpha^{(j)}_{m,m'}&=
{i\over2}(-1)^{m+m'}\sec\zeta\sum_{m''=1}^M {\rm sgn}(m''-m)
{\rm sgn}(m''-m')\left(e^{2i\zeta_{m''}}-1\right)\cr
&
\times\exp\left[i\sum_{m'''=1}^M
{\rm sgn}(m'''-m){\rm sgn}(m'''-m'')
{\rm sgn}(m''-m)\zeta_{m'''}\right].&(9.54)\cr}$$
We have used the abbreviations
$$\zeta_{m_j}={e\Delta\over2}A^j_{\overline{m_j-1}},\quad
\zeta=\sum_{m_j=1}^M\zeta_{m_j},\eqno(9.55)$$
and
$${\rm sgn}(x)=\cases{+1,\, &$x>0$,\cr
-1,\, &$x\le0$.\cr}\eqno(9.56)$$
We can now carry out the sum over $m''$ in (9.54):
$$\eqalignno{\alpha^{(j)}_{m,m'}&=i\epsilon_{m',m}(-1)^{m+m'}\left[
-1+\cos\left(\sum_{m''=1}^M
{\rm sgn}(m''-m){\rm sgn}(m''-m')\zeta_{m''}\right)\sec\zeta
\right]\cr
&-(-1)^{m+m'}\sin\left(\sum_{m''=1}^M
{\rm sgn}(m''-m){\rm sgn}(m''-m')\zeta_{m''}\right)
\sec\zeta,&(9.57)\cr}$$
where
$$\epsilon_{m',m}=
\cases{1, &$m'>m$,\cr
0, &$m'=m$,\cr
-1, &$m'<m$.\cr}\eqno(9.58)$$
It is obvious that $\alpha^{(j)}$ is Hermitian.

Let us write the Dirac equation (9.53) in the form
$$ U\psi_{n+1}=V\psi_n.\eqno(9.59)$$
It is apparent that $V=2-U$, so the transfer matrix is
$$ T=2 U^{-1}-1.\eqno(9.60)$$
The condition that $T$ is unitary translates into the
following condition on $U$:
$$U+U^\dagger=2.\eqno(9.61)$$
{}From (9.53) the matrix $U$ is explicitly
$$\eqalignno{
U_{{\bf m},{\bf m}'}&=\delta_{{\bf m},{\bf m}'}
+{h\over \Delta}
\gamma^0\gamma^j(-1)^{m_j+m_j'}\epsilon_{m_j,m_j'}\delta_{{\bf
m}_\perp,
{\bf m}_\perp'}\cr
&-{ih\mu\gamma^0\over2}\delta_{{\bf m},{\bf
m}'}-{ih\gamma^0\gamma^j\over\Delta}
\alpha^{(j)}_{{{\bf m}_\perp};
m_j,m_j'}\delta_{{\bf m}_\perp,{\bf m}_\perp'}.&(9.62)}$$
Therefore, the unitarity condition (9.61) is equivalent
to the condition that $\alpha^{(j)}$ be Hermitian:
$$ \alpha^{(j)\dagger}=\alpha^{(j)},\eqno(9.63)$$
which  is satisfied as noted above.
\bigskip
\noindent{\bf F.~Solution of the Abelian Equations---The Schwinger
Model}
\medskip

Here we wish to illustrate how the equations of motion are solved,
and
spectral information extracted, for the special case of the Schwinger
model, electrodynamics in two space-time dimensions with  zero bare
fermion mass.  We can work in
the gauge $A_0=0$, and choose a square lattice, $\Delta =h$, in which
case the
difference equation (9.34) can be solved by
$$\eqalignno{ \phi_{m,n+1}^{(+)}=
&e^{iehC_{m,n}}\phi_{m-1,n}^{(+)},&(\hbox{9.64a})\cr
\phi_{m-1,n+1}^{(-)}=&e^{-iehC_{m,n}}
\phi_{m,n}^{(-)},&(\hbox{9.64b})\cr}$$
where the superscripts denote the chirality, the eigenvalues of
$i\gamma_5
=\gamma^0\gamma^1$.  We take $\phi_{m,n}^{(\pm)}$ to be the canonical
fermion variables, which for free fields have the Fock--space
expansion
$$\phi_{m,n}=\sum_{k=1}^M e^{ip_kmh}(e^{-ip_knh}v^{(+)}a_k^{(+)}
+e^{ip_knh}v^{(-)}a_k^{(-)}),\eqno(\hbox{9.65})$$
where for $M$ even, $p_k =(2k+1)\pi/(Mh)$, $v^{(\pm)}$ are the
eigenvectors of
$\gamma_5$, and
$$[a_k^{(\pm)}, a_{k^\prime}^{(\pm)\dag}]_+={1\over Mh}\delta_{k
k^{\prime}}, \eqno(\hbox{9.66})$$
the other anticommutators being zero.  The physical interpretation of
$a^{(\pm)}_k$, $a^{(\pm)\dag}_k$ as creation and annihilation
operators is
as follows:
$$\eqalign{ \hbox{ for}\quad 0\le k\le {M\over2}-1,\qquad
&a_k^{(+)}|0\rangle
=a_k^{(-)\dag}|0\rangle=0,\cr \hbox{ for}\quad -{ M\over2}\le
k<0,\qquad &
a_k^{(+)\dag}|0\rangle=a_k^{(-)}|0\rangle=0;\cr}\eqno(\hbox{9.67})$$
this construction then implies the correct lattice fermion Green's
function.

The only physical particle in the Schwinger model is a boson, which
we denote
by $B$, of mass $\mu$.  We can obtain$^{29}$ the dispersion relation
for this
particle in a manner analogous to that employed in Sec.~VII by taking
matrix
elements of the equations of motion between the vacuum and a $B$
state of
momentum $q_l=2\pi l/Mh$. Using (9.15) we find
$$\eqalignno{\langle B,l|J_{m,n}^{(+)}|0\rangle=&{1\over2}\langle
B,l|J^0_{m,n}+J^1_{m,n}|0
\rangle\cr=&{1\over4h}\{-(e^{i\omega_lh}+1)
(e^{-iq_lh}-1)+(e^{i\omega_lh}-1)
(e^{-iq_lh}+1)\}\langle B,l|E_{m,n}|0\rangle\cr
\approx& {i\over2}(q_l+\omega_l)\langle
B,l|E_{m,n}|0\rangle,&(\hbox{9.68})\cr}$$
as $h\to 0$. On the other hand, we evaluate the current matrix
element using
the solution of the Dirac equation (9.64) and the Fock--space
expansion at the
initial time $n=0$.  That is, since
$$\phi_{m,n+1}^{(+)}=e^{ieh\sum_{r=0}^n C_{m-n+r,r}}\phi_{m-n-
1,0}^{(+)},\eqno(\hbox{9.69})$$
and
$$\phi_{m,0}^{(+)}=\sum_{k=1}^M
e^{ip_kmh}a_k^{(+)},\eqno(\hbox{9.70})$$
we have
$$\eqalignno{ J_{m,n}^{(+)}=& {e\over4}\sum_{k,k^\prime=1}^M
a_k^{(+)\dagger}
e^{i(p_{k^\prime}-p_k)(m-n)h}[1+e^{i(p_k-p_{k^\prime})h}\cr
&\quad+ e^{ip_kh}e^{-ieh\sum_{r=0}^n (C_{m-n+r,r}-C_{m-
n+1+r,r})}e^{-iehC_{m+1,n}}\cr
&\quad+ e^{-ip_{k^\prime}h}e^{ieh\sum_{r=0}^n
(C_{m-n+r,r}-C_{m-n+1+r,r})}
e^{iehC_{m+1,n}}]a_{k^\prime}^{(+)}.&(\hbox{9.71})\cr}$$
We now assume that the $B$ states are {\it not} created by fermion
operators,
that is, for all $k>0$,
$$\langle B,l|a_k^{(+)\dag} =0, \eqno(\hbox{9.72})$$
and that the commutator of $a_k^{(+)}$ and $C_{m,n}$ is negligible as
$h\to 0$.
Then, using the canonical relation (9.66) together with the vacuum
definition
(9.67), we find that the matrix element of (9.71) is
$$\eqalignno{\langle B,l|J_{m,n}^{(+)}|0\rangle=&
{e\over4}{1\over Mh}ieh\sum_{k=-M/2}^{-1}
(e^{-ip_kh}-e^{ip_kh})\cr &\quad\times\langle B,l|C_{m+1,n}
+\sum_{r=0}^n(C_{m-n+r,r}-C_{m-n+1+r,r})|0\rangle,
&(\hbox{9.73})\cr}$$
where an expansion in $h$ has been carried out.  The sum on $k$ in
(9.73) is
immediately evaluated as $2i/\sin(\pi/M)$, while the remaining matrix
element is
$$\langle
B,l|C_{m,n}|0\rangle\left[e^{-iq_lh}+iq_lh\sum_{r=0}^ne^{ih(q_l
-\omega_l)r}\right]\approx{\omega_l\over\omega_l-q_l}\langle
B,l|C_{m,n}|0\rangle, \eqno(\hbox{9.74})$$
where in the last summation on $r$ we have deleted a rapidly
oscillating term
$\sim e^{-i(\omega_l-q_l)nh}$ $\to0$ ($nh\to\infty$).  Finally, from
(9.12) we
learn, as $h\to 0$, that
$$i\omega_l\langle B,l|C_{m,n}|0\rangle=\langle B,l|E_{m,n}|0\rangle.
\eqno(\hbox{9.75})$$
When we put (9.68), (9.73), (9.74), and (9.75) together, we obtain
the desired
dispersion relation,
$$\eqalignno{ \langle B,l|E_{m,n}|0\rangle =&{2\over
i}{1\over\omega_l+q_l}
\langle B,l|J_{m,n}^{(+)}|0\rangle\cr
=&{2\over i}{1\over\omega_l+q_l}{ie^2\over4M}{2i\over\sin\pi/M}
{\omega_l\over\omega_l-q_l}\langle B,l|C_{m,n}|0\rangle\cr
=&{1\over\omega_l^2-q_l^2}{e^2\over M\sin\pi/M}\langle
B,l|E_{m,n}|0\rangle,
&(\hbox{9.76})\cr}$$
or
$$\omega_l^2=q_l^2+\mu^2,\eqno(\hbox{9.77})$$ where the mass $\mu$ is
$$\mu^2={e^2\over M\sin\pi/M}.\eqno(\hbox{9.78})$$

Using the solution (9.64) we may also calculate the divergences of
the
vector and axial currents. It is easily seen from (9.64) that the
lattice divergence of  the vector current
$$J^\mu_{m,n}=e\Psi_{m,n}^\dagger\gamma^0\gamma^\mu\Psi_{m,n}
\eqno(9.79)$$
is
$$(\hbox{``}\partial^\mu J_\mu\hbox{''})_{m,n}={ie\over2
h}\sin\left({e h^2\over2}\tilde E_{m,n}
\right)e^{ieh(C_{m+1,n+1}+C_{m,n+1})/2}
\phi^\dagger_{m+1,n+1}\phi_{m,n+1}+h.c., \eqno(9.80)$$
which has vanishing vacuum expectation value because
$$\langle\phi_{m,n}\phi^\dagger_{m+1,n}\rangle=-{i\gamma_5\over M
h\sin(\pi/M)}.
\eqno(9.81)$$
On the other hand, this same result shows that the axial-vector
current
$$J_5^\mu{}_{m,n}=e\Psi_{m,n}^\dagger\gamma^0\gamma^\mu\gamma_5
\Psi_{m,n}\eqno(9.82)$$ yields the axial anomaly
$$\langle(\hbox{``}\partial_\mu J_5^\mu\hbox{''})_{m,n}\rangle =
-e^2\left(M\sin{\pi\over M} \right)^{-1}\tilde E_{m,n}.\eqno(9.83)$$
The same anomaly emerges as in (9.78). The details of this latter
calculation
can be found in Ref.~27. Note that the lattice anomaly
$e^2/(M\sin(\pi/M))$
differs from the continuum value $e^2/\pi$ by a term of order
$1/M^2$, an
error typical of the linear finite-element approximation.
\vfill\eject
\bigskip
\noindent{\bf X.~LATTICE YANG-MILLS EQUATIONS}
\bigskip

We now continue the discussion where we left off in Sec.~IX.C. For
simplicity,
we continue to work in $1+1$ dimensions, and we will here assume a
square
lattice, $h=\Delta$.
\bigskip
\noindent{\bf A.~Construction of the Field Strength}
\medskip

Now that the gauge transformations for $C$ and $B$ are completely
determined, it should be straightforward to work out the interaction
terms in the field strength, $E=F_{01}$.
We begin by recalling that the zeroth order construction (9.12)
$$\tilde E^{(0)}_{m,n}={1\over h}(C_{m+1,n+1}-C_{m+1,n})-{1\over
h}
(B_{m+1,n+1}-B_{m,n+1}),\eqno(10.1)$$
is invariant under (9.23),
$$\eqalignno{\de^{(0)}C_{m,n}&
={1\over h}(\de\om_{m,n}-\de\om_{m-1,n}),&(10.2\hbox{a})\cr
\de^{(0)}B_{m,n}&={1\over
h}(\de\om_{m,n}-\de\om_{m,n-1}).&(10.2\hbox{b})\cr}$$
However, under the $\de^{(1)}$ transformations (9.28) and (9.44) the
variation of $\tilde E^{(0)}$ is
$$\eqalignno{\de^{(1)}\tilde E^{(0)}_{m,n}&=ig[\de\om_{m,n},\tilde
E^{(0)}
_{m,n}]\cr &\quad+{ig\over2}\delta^{(0)}\{[B_{m+1,n+1},C_{m+1,n+1}]
+[B_{m,n+1},C_{m+1,n+1}]+[B_{m+1,n+1},C_{m+1,n}]\cr
&\quad\quad-[B_{m,n+1},C_{m+1,n}]+[C_{m+1,n},C_{m+1,n+1}]-[B_{m,n+1},
B_{m+1,n+1}]\}.&(10.3)\cr}$$
The total variation in (10.3) allows us to identify the first-order
interaction term. Because it is already clear that the interaction
terms in
the field strength will be local, involving fields at the four
corners of the
finite element, let us simplify the following by introducing the
abbreviations $$\eqalignno{B_1\equiv B_{m+1,n+1},\quad\hbox{and}&
\quad B_0\equiv B_{m,n+1},&(10.4\hbox{a})\cr
C_1\equiv C_{m+1,n+1},\quad\hbox{and}&\quad C_0\equiv C_{m+1,n}.&
(10.4\hbox{b})\cr}$$
Then, the free term (10.1) is
$$\tilde E^{(0)}_{m,n}={1\over h}(C_1-C_0-B_1+B_0),\eqno(10.5)$$
and the interaction deduced from (10.3) is
$$\tilde
E^{(1)}_{m,n}=-{ig\over2}\{[B_1,C_1]+[B_0,C_1]+[B_1,C_0]-[B_0,C_0]
+[C_0,C_1]-[B_0,B_1]\}.\eqno(10.6)$$
Note that $E$ is antisymmetric under the interchange
$B\leftrightarrow C$,
an antisymmetry that will be maintained in each order.
We could add here an arbitrary multiple of $(\tilde E^{(0)}_{m,n})^2$
because
this is invariant under $\de^{(0)}$ variations. We choose not to do
so,
and thereby keep a minimal form in terms of nested commutators.
This form, of course, guarantees hermiticity.

It is straightforward to find the higher-order interaction terms.
We find $\tilde E^{(2)}$ from the equation
$$\de^{(2)}\tilde E^{(0)}_{m,n}+\de^{(1)}\tilde E^{(1)}_{m,n}
+\de^{(0)}\tilde E^{(2)}_{m,n}
=ig[\de\om_{m,n},\tilde E^{(1)}_{m,n}].\eqno(10.7)$$
Up to the above-mentioned ambiguity it is
$$\eqalignno{\tilde E^{(2)}_{m,n}
&=-{g^2 h\over12}\{[B_1,[B_1,C_1]]-[C_1,[C_1,B_1]
]-[B_1,[B_1,C_0]]+[C_1,[C_1,B_0]]\cr
&\quad\quad+2[B_1,[B_0,C_1]]-2[C_1,[C_0,B_1]]
+2[B_0,[B_1,C_1]]-2[C_0,[C_1,B_1]]
\cr&\quad\quad-2[B_1,[B_0,C_0]]+2[C_1,[C_0,B_0]]
+4[B_0,[B_1,C_0]]-4[C_0,
[C_1,B_0]]\cr&\quad\quad+[B_0,[B_0,C_1]]-[C_0,[C_0,B_1]]
-[B_0,[B_0,C_0]]
+[C_0,[C_0,B_0]]\cr&\quad\quad-[B_0,[B_0,B_1]]+[C_0,[C_0,C_1]]
-[B_1,[B_0,B_1]]+[C_1,[C_0,C_1]]\}.&(10.8)\cr}$$
Our final explicit example is $\tilde E^{(3)}$, obtained from
$$\de^{(0)}\tilde E^{(3)}_{m,n}+\de^{(1)}\tilde E^{(2)}_{m,n}
+\de^{(2)}\tilde E^{(1)}_{m,n}
=ig[\de\om_{m,n},\tilde E^{(2)}_{m,n}].\eqno(10.9)$$
(Recall that $\de^{(3)}=0$.)  It has fewer terms than might have been
anticipated:
$$\eqalignno{\tilde E^{(3)}_{m,n}&={ig^3 h^2\over
24}\{[B_1,[B_0,[B_0,B_1]]]
-[C_1,[C_0,[C_0,C_1]]]+2[B_0,[C_0,[B_0,B_1]]]\cr
&\quad\quad-2[C_0,[B_0,[C_0,C_1]]]-2[B_1,[C_0,[B_0,B_1]]]
+2[C_0,[B_1,[B_0,B_1]]]
\cr&\quad\quad+2[C_1,[B_0,[C_0,C_1]]]
-2[B_0,[C_1,[C_0,C_1]]]+2[B_1,[B_0,[C_1,B_1
]]]\cr&\quad\quad-2[C_1,[C_0,[B_1,C_1]]]-2[B_0,[B_1,[B_0,C_1]]]
+2[C_0,[C_1,[C_0,B_1]]]\cr
&\quad\quad-[B_1,[C_1,[B_1,C_1]]]-[B_0,[C_0,[B_0,C_0]]]
-2[B_0,[C_1,[B_1,C_1]]]
\cr&\quad\quad+2[C_1,[B_0,[B_1,C_1]]]+2[C_0,[B_1,[C_1,B_1]]]
-2[B_1,[C_0,[C_1,B_1]]]\cr
&\quad\quad-2[B_0,[C_0,[B_0,C_1]]]+2[C_0,[B_0,[C_0,B_1]]]
+[B_1,[C_0,[B_1,
C_0]]]\cr&\quad\quad-[C_1,[B_0,[C_1,B_0]]]-2[B_0,[C_1,[B_1,C_0]]]
+2[C_1,[B_0,[B_1,C_0]]]\cr
&\quad\quad-2[C_1,[C_0,[B_1,B_0]]]+2[C_0,[C_1,[B_1,B_0]]]-2[B_0,[C_0,
[B_1,C_1]]]\cr&\quad\quad+2[C_0,[B_0,[C_1,B_1]]]\}&(10.10)\cr}$$

Although we can compute $\tilde E^{(N)}$ to any required order $N$,
we
have not yet discovered a general iterative formula for $\tilde
E^{(N)}$.
A notable feature is the appearance in $\tilde E^{(N)}$ of the term
$$(igh)^N{{\cal B}_N\over2}\sum_{i=1}^4[A_i,[A_i,[\dots,[A_i,\tilde
E^{(0)}
]\dots]]],\eqno(10.11)$$
where $\{A_i\}=\{B_0,B_1,C_0,C_1\},$ which, for example, results in a
significant simplification of (10.8), and explains the absence of
many terms
in (10.10), because ${\cal B}_3=0$. The calculational methods so far
developed
for extracting information about matrix elements
require the knowledge of only the first few orders in any case, so
the
first few $E^{(N)}$ should be sufficient for at least the initial
stages of the
finite-element solution.

The generalization to four dimensions is nontrivial but
straightforward.
Of course there are more Yang-Mills field components, and new
structures
emerge. The lowest-order interaction terms are given in Ref.~30.
\bigskip
\noindent{\bf B.~Construction of Yang-Mills Equations}
\medskip

We finally must construct the lattice analogue of the continuum
Yang-Mills
equations, (9.9).  Upon a moment's reflection, however, it is clear
that
the structure of the interactions in the Dirac equation carries over
to the
Yang-Mills equations, with multiplication by powers of the potentials
replaced by nested commutators.  The reason for this is as follows.
We must
solve the covariance equation
$$\de\tilde E_{m,n}=ig[\de\om_{m,n},\tilde E_{m,n}]\eqno(10.12)$$
for $\delta F$ and $\de E$, where $F$ and $G$ are given by (9.16).
Because necessarily the boson fields must be periodic, to do this
we require that the number of spatial lattice sites, $M$, be odd.
Then we find
$$\delta
F_{m,n}=ig[\de\om_{m,n},F_{m,n}]+{ig\over2}\sum_{m'=m+1}^{m+M}
(-1)^{m+m'}[\de\om_{m',n}-\de\om_{m'-1,n},F_{m',n}],\eqno(10.13)$$
and
$$\de G_{m,n}=ig[\de\om_{m,n},G_{m,n}]-ig\mathop{{\sum}'}_{n'=0}^n
(-1)^{n+n'}[\de\om_{m,n'}-\de\om_{m,n'-1},G_{m,n'}].\eqno(10.14)$$
These have just the form as the transformation of $\th$ and $\phi$,
(9.19) and (9.39), respectively.  Here, as in (9.37), we have adopted
the initial conditions, at fixed spatial coordinate $m$,
$$\de G_0={ig\over2}[\de\om_0+\de\om_{-1},G_0].\eqno(10.15)$$

Without more ado, we transcribe the form of the interaction terms.
For $F$ we suppress the local variable $n$ and write,
for the term to be added to the left-hand side of (9.15a),
$${\cal I}^{(N)}_m={1\over2}\sum_{m'=m+1}^{m+M}(-1)^{m+m'}
\sum_{k=1}^N{(-igh)^k\over k!}\underbrace{[C_{m'},[C_{m'},
[\dots,[C_{m'},{\cal I}_{m'}^{(N-k)}]\dots]]]}
_{k{\rm\;nested\; commutators}},\eqno(10.16)$$
and for $G$ we suppress the local variable $m$ and write,
for the term to be added to the left-hand side of (9.15b),
$$\eqalignno{{\cal K}^{(N)}_n&=-\sum_{n'=1}^n(-1)^{n+n'}\sum_{k=1}^N
{(-igh)^k\over k!}\underbrace{
[B_{n'},[B_{n'},[\dots,[B_{n'},{\cal K}_{n'}^{(N-k)}]
\dots]]]}_{k{\rm\; nested\; commutators}}
\cr&\quad+(igh)^N{1\over
2^{N}N!}(-1)^n\underbrace{[B_0,[B_0,[\dots,[B_0,
{\cal K}^{(0)}_0]
\dots]]]}_{N{\rm\; nested\; commutators}}.&(10.17)\cr}$$
Here ${\cal I}^{(0)}_m=-2F_m/h$ and ${\cal K}^{(0)}_n=-2 G_n/h$.
In Ref.~29 we prove the required covariance statements:
$$\de_F{\cal I}^{(N)}_m+\sum_{k=0}^N\delta^{(k)}{\cal I}^{(N-k+1)}_m
=ig[\de\om_m,{\cal I}^{(N)}_m],\eqno(10.18)$$
and $$\de_G{\cal K}^{(N)}_n+\sum_{k=0}^N\de^{(k)}{\cal K}^{(N-k+1)}_n
=ig[\de\om_n,{\cal K}^{(N)}_n].\eqno(10.19)$$
The Yang-Mills equations are given by (9.15), (10.16), and (10.17):
$$\eqalignno{{1\over h}(F_{m+1,n}-F_{m,n})+{\cal I}_{m,n}
&=-\tilde\jmath_{m,n}^0,&(10.20\hbox{a})\cr
{1\over h}(G_{m,n+1}-G_{m,n})+{\cal K}_{m,n}&=
\tilde\jmath_{m,n}^1,&(10.20\hbox{b})\cr}$$
where $$\eqalignno{{\cal I}_{m,n}&=\sum_{N=1}^\infty{\cal
I}^{(N)}_{m,n},
&(10.21\hbox{a})\cr
{\cal K}_{m,n}&=\sum_{N=1}^\infty{\cal
K}^{(N)}_{m,n}.&(10.21\hbox{b})\cr}$$

Here, we easily derive from (10.16) and (10.17) first
$$\eqalignno{{\cal
I}_m&={1\over2}\sum_{m'=m+1}^{m+M}(-1)^{m+m'}\left[
e^{-ighC_{m'}}\left({\cal I}^{(0)}_{m'}+{\cal
I}_{m'}\right)e^{ighC_{m'}}
-{\cal I}^{(0)}_{m'}-{\cal I}_{m'}\right],&(10.22\hbox{a})\cr
\noalign{and} {\cal
K}_n&=-\sum_{n'=1}^n(-1)^{n+n'}\left[e^{-ighB_{n'}}\left(
{\cal K}^{(0)}_{n'}+{\cal K}_{n'}\right)e^{ighB_{n'}}-{\cal
K}^{(0)}_{n'}
-{\cal K}_{n'}\right]\cr
&\qquad+(-1)^n\left[e^{ighB_0/2}{\cal K}^{(0)}_0e^{-ighB_0/2}-{\cal
K}_0
^{(0)}\right],&(10.22\hbox{b})\cr}$$
and then the difference equations
$$\eqalignno{{\cal I}_m e^{ighC_m}+e^{ighC_m}{\cal I}_{m-1}&=
-{2\over h}\left[e^{ighC_m},F_m\right],&(10.23\hbox{a})\cr
\noalign{and} {\cal K}_n e^{ighB_n}+e^{ighB_n}{\cal
K}_{n-1}&=-{2\over h}
\left[e^{ighB_n},G_n\right].&(10.23\hbox{b})\cr}$$

We are currently studying the solutions of these equation in two and
four space-time dimensions.
\bigskip
\bigskip

We thank G. Dunne for many extended and useful conversations.
LRM thanks the Physics Department at Washington University for its
hospitality.
We thank the U. S. Department of Energy for funding this research.
\vfill\eject

\centerline{\bf REFERENCES}
\bigskip
\item{$^1$} Strictly speaking, we are using the {\sl collocation}
method, a
technique closely resembling the method of finite elements.
\medskip
\item{$^2$} If one imposes (14a) at one point on the finite element,
say
$t=ah$, and (14b) at another point on the finite element, say $t=bh$,
then
the unitarity condition (15) is satisfied for all $n$ if $a+b=1$.
However, the
resulting difference scheme is only correct to order $1/N$ and not to
order
$1/N^2$.
\medskip
\item{$^3$} The inverse function $g^{-1}$ is unique if $V''(x)$ is
positive
(that is, if $V$ is a single potential well). The case of double
wells for
which $g^{-1}$ is not unique is an extremely interesting one
deserving
further investigation.
\medskip
\item{$^{4}$} C. M. Bender, L. M. Simmons, Jr., and R. Stong, Phys.
Rev. D
{\bf 33}, 2362 (1986).
\medskip
\item{$^{5}$} C. M. Bender, K. A. Milton, D. H. Sharp, L. M. Simmons,
Jr.,
and R. Stong, Phys. Rev. D {\bf 32}, 1476 (1985).
An alternative approach to quantum mechanics in which one attempts to
solve
the operator equations of motion in the continuum is considered in
C. M. Bender and G. V. Dunne, Phys.~Lett.~B {\bf 200}, 520 (1988);
Phys.~Rev.~D {\bf 40}, 2739 (1989); Phys.~Rev.~D {\bf 40}, 3504
(1989).
\medskip
\item{$^{6}$} C. M. Bender and M. L. Green, Phys. Rev. D {\bf 34},
3255
(1986).
\medskip
\item{$^{7}$} C. M. Bender, F. Cooper, J. E. O'Dell, and L. M.
Simmons, Jr.,
Phys. Rev. Lett. {\bf 55}, 901 (1985); C. M. Bender, F. Cooper, V. P.
Gutschick,
and M. M. Nieto, Phys.~Rev.~D {\bf 32}, 1486 (1985).
\medskip
\item{$^{8}$} C. M. Bender, L. R. Mead, and S. S. Pinsky,
Phys.~Rev.~Lett.
{\bf 56}, 2445 (1986).
\medskip
\item{$^{9}$} The polynomials described here are special cases of
continuous
Hahn polynomials of imaginary argument. These polynomials were
recently
discussed by N. M. Atakishiyev and S. K. Suslov, J. Phys. A {\bf 18},
1583
(1985), and R. Askey, J. Phys. A {\bf 18}, L1017 (1985).
\medskip
\item{$^{10}$} C. M. Bender, L. R. Mead, and S. S. Pinsky, J. Math.
Phys.
{\bf 28}, 509 (1987).
\medskip
\item{$^{11}$} There is an exact one-to-one correspondence between
all
possible sets of polynomials (both orthogonal and nonorthogonal) and
rules
for operator orderings. See C. M. Bender and G. V. Dunne, J. Math.
Phys.
{\bf 29}, 1727 (1988).
\medskip
\item{$^{12}$} This definition of $S_n (x)$ in (3.21) very closely
resembles
that of the Chebyshev polynomials $T_n (x)$. Using the fact that
$\cos
(n\theta)$ is a polynomial in $\cos (\theta )$ one defines $T_n (\cos
\theta
) \equiv \cos (n \theta )$.
\medskip
\item{$^{13}$} Equation (3.21) can be generalized to include
off-diagonal
$T$ forms:
$$T_{m, m+k} = {{(2m+k)!}\over {(m+k)! 2^{m+1}}} \left \{ q^k , S_m
(T_{1,1}) \right \} _+ .$$
\medskip
\item{$^{14}$} There is an interesting connection between these
polynomials
and the Euler numbers $E_n$: $\int_{-\infty}^{\infty}dx\; w(x)
x^{2n}=|E_{2n}|$.
\medskip
\item{$^{15}$} P. M. Prenter, {\sl Splines and Variational Methods}
(Wiley,
New York, 1975).
\medskip
\item{$^{16}$} L. Durand, Proc.~Int.~Symp. on
Orthogonal Polynomials and Their Applications, Bar-le-Duc, France,
1984.
\medskip
\item{$^{17}$} C. M. Bender, F. Cooper, K. A. Milton, S. S. Pinsky,
L. M. Simmons, Jr., Phys. Rev. D {\bf 35}, 3081 (1987).
\medskip
\item{$^{18}$} C. M. Bender, K. A. Milton, S. S. Pinsky, and L. M.
Simmons, Jr.,
Phys.~Rev.~D {\bf 33}, 1692 (1986).
\medskip
\item{$^{19}$} C. M. Bender and D. H. Sharp, Phys.~Rev.~Lett.~{\bf
50}, 1535
(1983). For an alternative application of finite elements to the
evaluation of
functional integrals in quantum field theory see C. M. Bender, G. S.
Guralnik,
and D. H. Sharp, Nucl.~Phys. {\bf B207}, 54 (1982).
\medskip
\item{$^{20}$} C. M. Bender and K. A. Milton, Phys.~Rev.~D {\bf 34},
3149
(1986).
\medskip
\item{$^{21}$} R. F. Dashen, B. Hasslacher, and A. Neveu,
Phys.~Rev.~D
{\bf 11}, 3424 (1975).
\medskip
\item{$^{22}$} S. Coleman, Phys.~Rev.~D {\bf 11}, 2088 (1975).
\medskip
\item{$^{23}$} C. M. Bender, K. A. Milton, and D. H. Sharp,
Phys.~Rev.~Lett.~{\bf51}, 1815 (1983).
\medskip
\item{$^{24}$} D. Miller, K. A. Milton, and S. Siegemund-Broka,
Phys.~Rev.~D
{\bf 46}, 806 (1993).
\medskip
\item{$^{25}$} Equation (8.8) was also discovered by R. Stacey; see
Phys.~Lett.
{\bf 129B}, 239 (1983) and Phys.~Rev.~D {\bf 26}, 468 (1982). The
finite-element
result was generalized to arbitrary dimension by T. Matsuyama (1984
preprint).
\medskip
\item{$^{26}$} L. H. Karsten and J. Smit, Nucl.~Phys.~{\bf B183}, 103
(1981);
H. B. Nielsen and M. Ninomiya, Nucl.~Phys.~{\bf B185}, 20 (1981);
J. M. Rabin, Phys.~Rev.~D {\bf 24}, 3218 (1981).
\medskip
\item{$^{27}$} C. M. Bender, K. A. Milton, and D. H. Sharp,
Phys.~Rev.~D {\bf 31}, 383 (1985).
\medskip
\item{$^{28}$} K. A. Milton and T. Grose, Phys.~Rev.~D {\bf 41}, 1261
(1990).
\medskip
\item{$^{29}$} T. Grose and K. A. Milton, Phys.~Rev.~D {\bf 37}, 1603
(1988).
\medskip
\item{$^{30}$} K. A. Milton, in {\it Proceeding of the XXVth
International Conference on High-Energy Physics}, Singapore, 1990,
edited by
K. K. Phua and Y. Yamaguchi (World Scientific, Singapore, 1991),
p.~432.
\medskip
\bye